# Revealing crucial effects of reservoir environment and hydrocarbon fractions on fluid behaviour in kaolinite pores

**Rixin Zhao[a, b, c], Haitao Xue[a, b], Shuangfang Lu[d, *], H. Chris Greenwell[c], Valentina Erastova[e, *]**

[a]School of Geosciences, China University of Petroleum (East China), Qingdao 266580, Shandong, China

[b]Key Laboratory of Deep Oil and Gas, China University of Petroleum (East China), Qingdao, 266580, China

[c]Department of Earth Sciences and 〃Department of Chemistry, Durham University, Durham DH1 3LE, United Kingdom

[d]Sanya Offshore Oil & Gas Research Institute, Northeast Petroleum University, Sanya 572025, China

[e]School of Chemistry, University of Edinburgh, David Brewster Road, Edinburgh, EH9 3FJ, United Kingdom

E-mail addresses: lushuangfang@upc.edu.cn (Shuangfang Lu), valentina.erastova@ed.ac.uk (Valentina Erastova).

## Abstract

The adsorption interactions of hydrocarbons and clay surfaces are crucial to understanding fluid behaviour within shale reservoirs and to mediating organic pollutants in soils. These interactions are affected by the diversity of complex hydrocarbon components and the variations in environmental conditions. This study examines the interactions between kaolinite clay, featuring two distinct basal surfaces, and an array of hydrocarbons. We assess the impact of various molecular structures, functional groups, and environmental conditions (focusing on the reservoir temperature and pressure ranges) on the adsorption selectivity, surface packing, molecular alignment and orientation, and diffusion of hydrocarbons. Analyses of molecular interaction energies provide a quantitative elucidation of the adsorption mechanisms of hydrocarbons on the different kaolinite surfaces. Our findings suggest that molecular configuration, functional group properties, and spatial effects dictate the distribution patterns of hydrocarbons for the different kaolinite surfaces. The differences in the interaction energy between various hydrocarbons with kaolinite reveal the adsorption strength of different hydrocarbons in the order of asphaltenes > heteroatomic hydrocarbons > saturated hydrocarbons > aromatic hydrocarbons. Furthermore, we observe that the adsorptive characteristics of hydrocarbons on kaolinite are highly temperature-sensitive, with increased temperatures markedly reducing the adsorption amount. Beyond a certain threshold, the effect of pressure rise on the fluid behaviour of hydrocarbons is non-negligible and is related to molecular packing and reduced mobility. Simulation results based on actual geological characteristics demonstrate notable adsorption disparities among hydrocarbon components on different kaolinite surfaces, influenced by competitive adsorption and clay surface interactions.

Polar surfaces are predominantly occupied by heteroatomic hydrocarbons, whereas on non-polar surfaces, asphaltenes and heavy saturated hydrocarbons develop multi-layer adsorption structures, with molecules aligned parallel to the surface.

## 1. Introduction

The increasing disparity between global energy demands and supply shortages underscores the strategic significance of unconventional oil reservoirs, regarded as crucial alternative resources, key to ensuring resource security due to their considerable reserves [1]. Following the triumphant exploration and development of North American shale oil and gas, igniting a global shale revolution, shale has transitioned from a conventional "source-seal" system to a "source-reservoir-seal" system. It has emerged as a focal point in the field of unconventional oil and gas, which has generated substantial interest in the academic and industrial sectors [2–4]. Decades of research, supported by numerous examples and data, have corroborated the existence of a multitude of micro- to nano-scale pores and fractures within shale formations [5,6]. In typical lacustrine basins, such as the Songliao Basin in China, the content of clay minerals such as illite, montmorillonite and kaolinite is notably high, reaching 40 %-60 % of the total mineral mass (Fig. S1 (a)). Recent studies show that nanoscale interparticle pores within clay minerals in organic-rich shales constitute the primary reservoir space networks [7]. Characterised by small particle size, large specific surface area (approximately 800 $m^2/g$), and heightened chemical reactivity at surface defect sites, clay minerals exhibit exceptional adsorption capacities for metal ions, protons, and organic molecules [8,9]. In contrast to montmorillonite, kaolinite is characterized by its stability and non-swelling characteristics, maintaining consistent basal surface charges regardless of surrounding solution proton concentration or pH alterations [10]. Consequently, kaolinite plays a significant role in the formation, migration, and accumulation of shale oil [11]. Additionally, kaolinite, other clay minerals and their modified composites have been extensively used for adsorbing heavy metals and organic pollutants from soil and aqueous environments [12–14]. Hence, delving into the adsorption mechanisms of kaolinite and clay on organic molecules, as well as their influencing factors, is imperative for elucidating fluid distribution patterns in shale reservoirs and effectively removing organic pollutants in environmental contexts.

In the context of shale oil production and environmental remediation processes, including oil sludge removal and organic pollutant extraction, the adsorption of hydrocarbons on clay surfaces represents a ubiquitous yet complex challenge [15,16]. Extensive research has been conducted to investigate fluid-solid interface interactions, including the adsorption behaviours of diverse organic hydrocarbon components on clay surfaces [17,18], the chemical influence of water and ions [19], and the impact of temperature [20,21], among other factors. These studies indicate that the specific characteristics of hydrocarbons and environmental conditions play significant roles in their interactions with clay surfaces. Nonetheless, the intricate interaction mechanisms between clay and various types of hydrocarbons have not been well elaborated, and our knowledge in modulating the adsorption and separation of organic hydrocarbon components on clay surfaces remains limited.

Shale oil is a complex mixture consisting of hydrocarbons (saturated and aromatic hydrocarbons) and nitrogen, sulphur, and oxygen-containing compounds (commonly referred to as asphaltenes and resins) [22]. Owing to the complexity of shale oil components, pronounced fractionation effects emerge during hydrocarbon expulsion, resulting in distinct disparities in the composition of families between retained and expelled hydrocarbons [23]. Revealing the interactions between complex oil components and clay under geological temperature and pressure conditions, and identifying the determinants and microscale interaction mechanisms of organic hydrocarbon adsorption and separation, are pivotal in comprehending fluid accumulation processes in shale reservoirs and evaluating their mobilization potential. Additionally, this also contributes to better optimizing the application of kaolinite across various fields, such as in organic pollutant absorption [24], as drug delivery carriers [25], and in enhancing oil recovery and drilling fluid innovation within the petroleum and natural gas industries [26]. Several pioneering experimental studies have hypothesized about the attributes and influential factors of organic hydrocarbon adsorption and separation, suggesting that hydrocarbons form regularly arranged adsorption layers near solid walls [27,28] and that adsorption amounts are sensitive to pore size and temperature [29]. However, quantitative microscopic evaluation is still lacking to consolidate these conjectures. The prevalent experimental techniques for studying the adsorption mechanism on clay, such as multi-step rock-eval pyrolysis [30], nuclear magnetic resonance methods [31–33], and microscopic observation methods [34,35], cannot capture the actual physicochemical properties of mineral surfaces, the variability in hydrocarbon components, and the impact of environmental conditions on adsorption behaviour, as these tests are restricted to static conditions and limited temperature and pressure ranges.

Molecular dynamics (MD) simulation methods are regarded as an effective bridge between experimental observations and microscopic mechanisms, attributed to their technical qualities [36]. These methods provide an alternative means for examining complex interactions between clay and hydrocarbons at the nanoscale, furnishing atomic-scale insights into their underlying mechanisms [37,38]. However, the diverse adsorption mechanisms of various organic hydrocarbon components on clay surfaces have been reported infrequently due to the difficulties in developing complex clay-organic hydrocarbon interaction models, while the studies on the impact of kaolinite surface characteristics on the fluid dynamics of hydrocarbon components are also sparse. Tunega et al. discerned contrasting hydrophilic and hydrophobic characteristics between the octahedral and tetrahedral surfaces of kaolinite, respectively [39]. Given that crude oil encompasses over 45% non-polar components and less than 15% polar components [40], it becomes imperative to investigate the divergent interaction and fluid behaviour mechanisms between different hydrocarbon components and kaolinite's two distinct basal surfaces. The limitations of the existing numerical simulation studies have been identified as: (i) the mineral-organic hydrocarbon interaction is predominantly analysed qualitatively through molecular configurations and density distributions, lacking quantitative methodologies for assessing microscopic-scale fluid behaviour; (ii) in explorations of the environmental temperature and pressure impacts on organic hydrocarbon adsorption, there is a lack of comparative discourse on atomic-scale interaction energies between hydrocarbons and clay. Despite the prevalent belief that pressure minimally affects organic hydrocarbon adsorption [41], recent experimental studies suggest an opposite result [42], implying that the combined influence of temperature and pressure on the fluid behaviour of hydrocarbons on clay may be profoundly complicated. Previous studies have rarely considered the changes in spatial effects and molecular orientation in simulation systems, thus not elucidating the microscopic mechanisms behind environmental variations influencing fluid adsorption behaviour; (iii) The differential adsorption of isolated monomeric hydrocarbons on clay and the competitive adsorption phenomena among mixed-component hydrocarbons remain inadequately explained [28,43].

In this work, we employ MD simulations to investigate the adsorption dynamics of hydrocarbons on kaolinite basal surfaces under geological temperature and pressure conditions, with a focus on elucidating the microscopic mechanisms involved, the findings of which will overcome the identified research limitations mentioned above. Our study combines kaolinite with two distinct surface properties and 15 types of organic hydrocarbon molecules, representing four primary components of crude oil. The main aims of this work are: (i) quantitatively assess the adsorption behaviour and interaction strengths between varying basal surfaces of kaolinite and diverse organic hydrocarbon components; (ii) deepen our understanding of the distribution patterns of hydrocarbon components on kaolinite surfaces based on molecular orientation features; (iii) explore the effects of temperature and pressure on the adsorption behaviour of hydrocarbons; (iv) elucidate the impact of competitive adsorption among mixed-component hydrocarbons on their adsorption behaviours.

# 2. Molecular models and methodology

## 2.1 Molecular models construction

The random contact between kaolinite and hydrocarbon molecules in natural settings results in multiple sites for mutual interactions, which can be categorized into edge surface sides, outer surface sites, interlayer sites, and inter-particle sites [44]. Structurally, kaolinite belongs to the 1:1 nonswelling group of clay minerals, where each layer of the clay consists of one tetrahedral silica sheet (T-$SiO_4$) and one octahedral alumina sheet (O-$AlO_6$). Unlike the 2:1 clays (e.g., illite and montmorillonite) with TOT structure, the OT structure of kaolinite results in distinct properties on each of the basal surfaces (see Fig. 2a). Specifically, the octahedral basal surface will be terminated with hydroxyl groups, leading to the common nomenclature of this surface as the hydroxyl surface (HY), while the tetrahedral surface is referred to as the siloxane surface (SI). Silica-based minerals, such as quartz, are widely employed for investigating fluid interactions and transport in inorganic matrices. Consequently, the hydroxyl surface of kaolinite is often modelled to represent the basal surface properties of silica-based minerals [45–48]. Therefore, selecting a sandwich-like structure consisting of hydrocarbon molecules encapsulated between two basal surfaces provides an insightful representation of the disparate interactions between kaolinite clay's distinct basal faces and organic molecules.

Owing to its low number of isomorphic substitutions, kaolinite does not hold a permanent charge and does not exhibit swelling behaviour. Instead, the charge on kaolinite is dependent on the pH of the surrounding solution, particularly owing to the deprotonation of hydroxyl groups on the octahedral surface. The basal surfaces of kaolinite exhibit significant differences in charge intensity under various pH conditions. Untreated and acid-treated kaolinite show pKa values similar to their aluminium hydroxide sites (5.49 and 5.94, respectively, close to 5.7), while the pKa value of alkali-treated kaolinite (6.34) is more aligned with the average pKa value of aluminium hydroxide and silicon hydroxide sites [49]. This indicates that the charge characteristics of kaolinite are closely related to the environmental pH value, with its net charge reaching zero at a pH of approximately 4.6 [50,51]. However, the observed phenomenon of charge reversal is primarily associated to the clay edges and is less prevalent at the basal surfaces [52]. Therefore, the present study focuses solely on the interactions of the two basal surfaces of the kaolinite with hydrocarbon molecules, without considering the edge sites or charge variations with pH. The dual basal surface structure of kaolinite is modelled based on the natural Georgia kaolinite (KGa-1b) features; the unit cell structure is shown in the SI Fig. S2, with initial atomic positions sourced from the American Mineralogist Crystal Structure Database [53,54]. The unit cell formula is given as:

$(Mg_{0.02}Ca_{0.01}Na_{0.01}K_{0.01})[Al_{3.86}Fe^{III}_{0.02}Mn_{tr}Ti_{0.11}][Si_{3.83}Al_{0.17}]O_{10}(OH)_8$. While the net charge of the unit cell is near-neutral, the individual atoms on the siloxane and hydroxyl surfaces carry slight negative and positive partial charges, respectively, as derived from quantum mechanical calculations [55,56]. These partial charges are implicitly captured through isomorphic substitution, which follows Loewenstein's rule [57]. The clay basal surfaces consist of 12×7 periodically replicated unit cells, producing a basal surface size of approximately 6.19×6.26 nm$^2$. Kaolinite is composed of a 3-layer stacked structure, and is positioned at the $0<z<1.92$ nm region of the supercell. Above this structural layer, an extra space of approximately 15 nm is integrated to formulate a pore region. This configuration is derived from prior research findings regarding the pore size distribution in shale (Fig. S1(c)).

***Table 1***

*Parameters and number of molecules per system of 16 organic hydrocarbon molecules.*

| Hydrocarbon Type | Chemical Structure | Formula | Molecular Weight (g/mol) | Loading Number of Molecules |
|---|---|---|---|---|
| Saturated |  n-Hexane | $C_6H_{14}$ | 86.180 | 2618 |
| |  n-Octane | $C_8H_{18}$ | 114.232 | 2102 |
| |  n-Dodecane | $C_{12}H_{26}$ | 170.330 | 1503 |
| |  n-Hexadecane | $C_{16}H_{34}$ | 226.448 | 1161 |
| |  n-Octadecane | $C_{18}H_{38}$ | 254.502 | 1042 |
| |  n-Eicosane | $C_{20}H_{42}$ | 282.556 | 953 |
| | 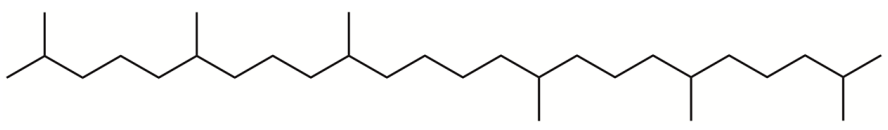 Squalane | $C_{30}H_{62}$ | 422.800 | 637 |
| Aromatic | 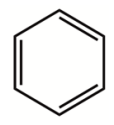 Benzene | $C_6H_6$ | 78.114 | 3828 |
| | 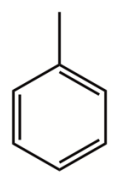 Toluene | $C_7H_8$ | 92.140 | 3196 |
| | 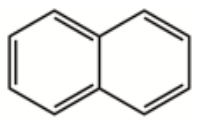 Naphthalene | $C_{10}H_8$ | 128.174 | 3096 |
| Heteroatomic | 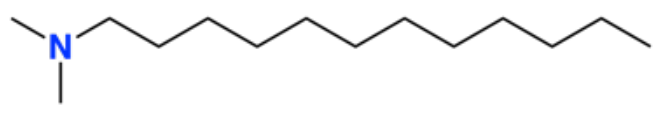  n,n-Dimethyldodecylamine | $C_{14}H_{31}N$ | 213.400 | 1243 |
| | 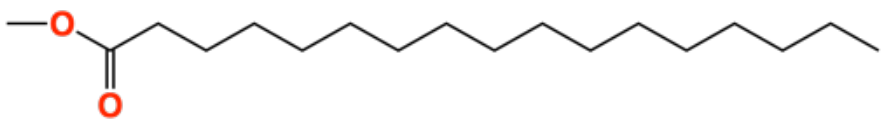 n-octadecanoic acid (Stearic Acid) | $C_{18}H_{36}O_2$ | 284.484 | 1129 |
| | 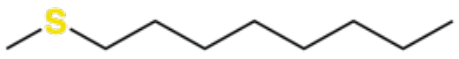 1-Octanethiol | $C_8H_{18}S$ | 146.300 | 1969 |
| Asphaltene | 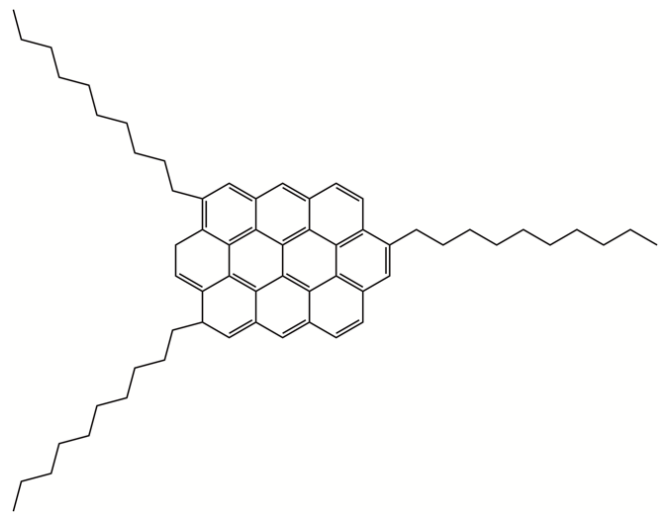 ASNP (Non-polar Asphaltene) | $C_{62}H_{85}$ | 847.383 | 403 |
| | 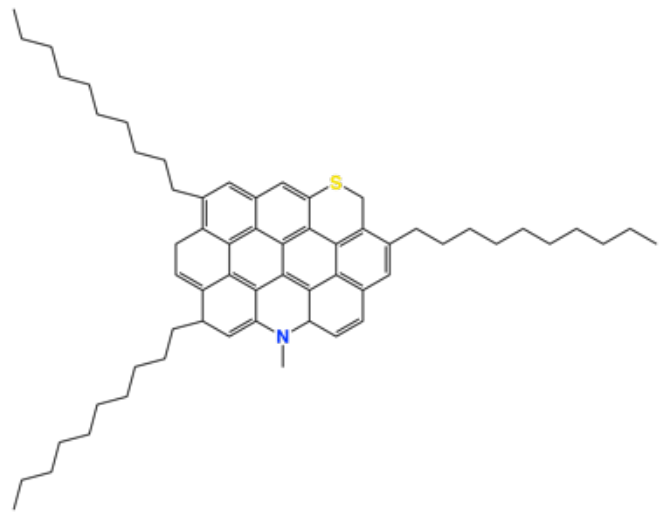 ASPO (Polar Asphaltene) | $C_{60}H_{81}NS$ | 830.362 | 411 |

To better elucidate the interaction characteristics of different types of hydrocarbon molecules on kaolinite basal surfaces, the present study selects 16 hydrocarbon molecules to represent the four key components of crude oil: saturated hydrocarbons, aromatic hydrocarbons, heteroatom-containing hydrocarbons, and asphaltenes (see Table 1). Alkanes with varying carbon chain lengths are chosen to represent non-polar saturated hydrocarbons, while squalene is selected to represent long-chain saturates with branching. Previous reports have identified squalane as a component of petroleum that can be extracted from both plant and animal sources [58]. Polar hydrocarbons can significantly alter the properties of hydrocarbon fluids in reservoirs, such as fluid viscosity, wettability, interfacial activity, and chemical stability [59,60]. The interactions between petroleum and rock matrices are mainly attributed to polar components, especially in hydrocarbons that usually retain more polar components [61]. Hydrocarbons containing nitrogen, oxygen, and sulfur are the major polar components in crude oil. As shown in Table 1, n,n-dimethyldodecylamine, stearic acid, and 1-octanethiol were chosen to represent heteroatomic hydrocarbons in crude oil [28,62]. Benzene, toluene, naphthalene, and dimethylnaphthalene are selected to represent aromatic hydrocarbons with and without branching. Li and Greenfield, building on the Mullins model, have established molecular structures that better reflect the properties of asphaltene [63,64]. In our work, we use two types of asphaltenes: one structure without heteroatoms, named ASNP, and another with nitrogen and sulphur heteroatoms, named ASPO (see Table 1). To understand how various hydrocarbon constituents and reservoir environmental conditions affect the molecular adsorption phenomena on kaolinite basal surfaces, we conducted two sets of molecular dynamics simulations. The first focused on the behaviour of isolated monomers under standard conditions (Fig. 2a).

Under actual geological reservoir conditions, the second set of studies focused at the adsorption of complex mixtures of hydrocarbons based on the actual ratios of crude oil components (Fig. 2b). Table 1 displays the type of natural hydrocarbon molecules that were chosen as well as the amount of molecules that were added into the mixed system. In the mixed system, the 29 hydrocarbons introduced were grouped into four categories, using the same classification method applied to isolated monomer hydrocarbon components. Their mass ratios are consistent with the results of the analysis of the composition of crude oil extracted from shales of different maturity levels in the Songliao Basin in China (vitrinite reflectance between 0.7 and 1, shown in SI Table S2). This approach representatively reflects the fluid behaviour of mixed hydrocarbons in clay pores under real geological conditions.

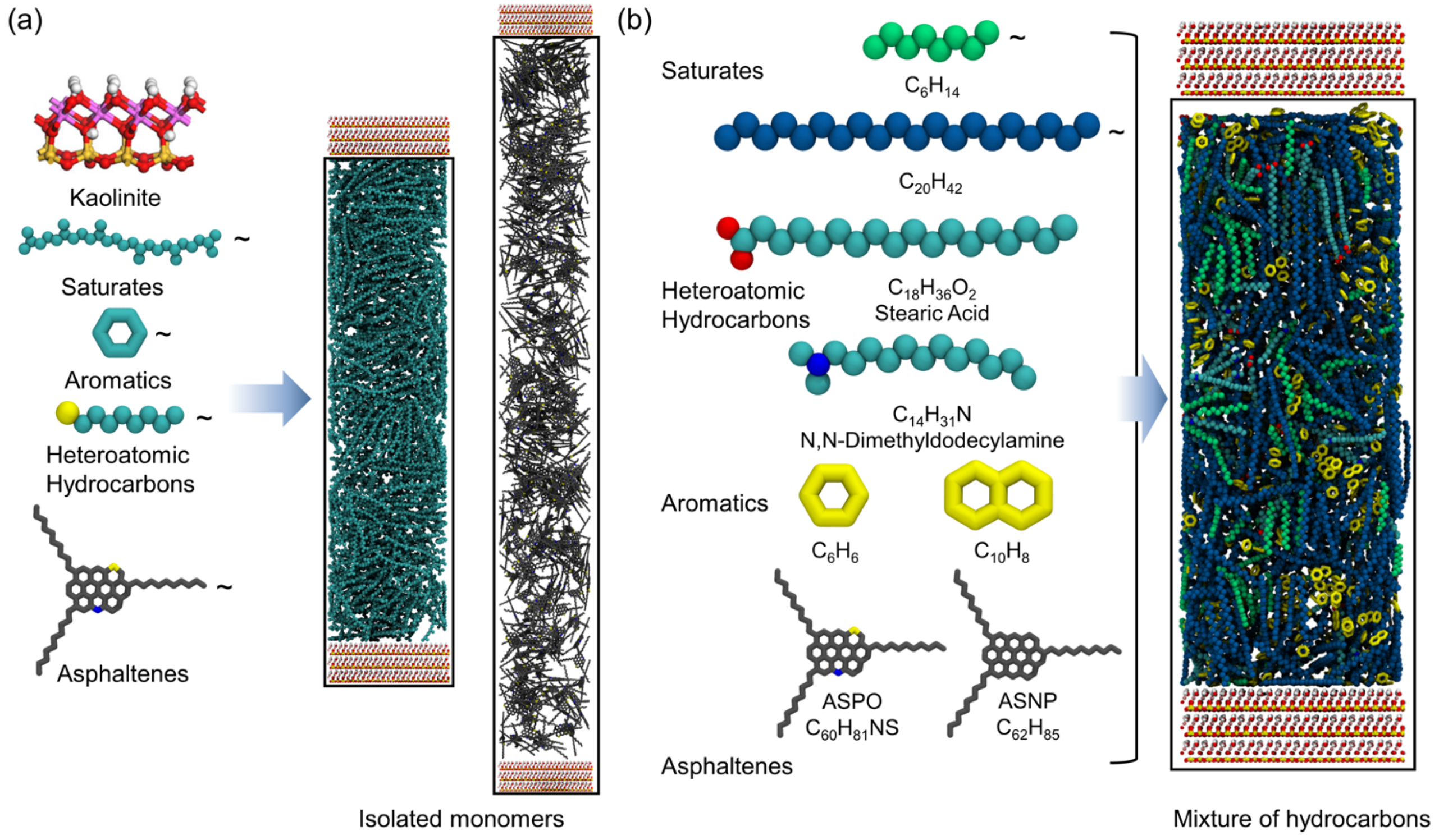

***Fig. 2.*** *(a) Schematic diagram of molecular models for isolated monomers and multi-component simulations on kaolinite basal surfaces depict molecular structures of individual monomer hydrocarbons and their initial models (specifically, squalene and asphaltene); (b) Schematic representation of diverse hydrocarbon molecules in the mixed system, wherein '~' symbolizes 'and so forth,' indicating that not all hydrocarbon molecules of identical components are illustrated.*

### 2.2 Choice of force fields

In this study, the ClayFF force field, parameterized specifically to simulate clay-like minerals and their interfacial properties [55], was employed to model kaolinite. The ClayFF force field, inclusive of a flexible single-point charge (SPC/E) model for water and hydroxyl groups, has been widely utilized for simulating mineral-organic interfaces [56,65,66]. Organic hydrocarbon molecules within the system were modelled using the CHARMM36 force field [67], which has been designed to accurately depict organic systems and has demonstrated reliability in physically representing hydrocarbons, lipids, and similar organic entities [68]. Both CHARMM36 and ClayFF employ the Lorentz–Berthelot mixing rules for simulating intermolecular interactions. The combination of ClayFF and CHARMM36 has been used in previous works and has been validated through joint tests to accurately reproduce the equilibrium properties of organic molecules interacting with hydrated mineral surfaces [69]. This has been verified by comparison with *ab initio* molecular dynamics results for the adsorption of acetate and ammonium molecules on quartz surfaces, as well as experimentally based X-ray diffraction data [70]. The total potential energy in the simulation was determined by summing up each energy contributing component as listed in SI Tables S3 and S4. Table S3 defines the potential energy terms for the kaolinite clay framework, while all the energy terms in Table S4 are used to describe organic hydrocarbon molecules. Detailed force field parameters are compiled in the Supplementary material (Table S5 and Table S6).

### 2.3 Molecular dynamics simulations

The initial models for all simulated systems were constructed via the Packmol 20.14.2 program [71] and optimized to eliminate anomalous molecular conformations, resulting in the model systems depicted in Fig. 2. Based on the density of various hydrocarbons at room temperature and the volume of the simulation box, the initial number of molecules to be loaded was calculated (see Table 1). Subsequently, all generated models were imported into the Gromacs 2020.2 suite for MD simulations [72]. All trajectory files and molecular configuration characteristics were visualised using VMD 1.9.4 [73]. Targeting the dispersion of overlapping or proximate atoms, energy minimization was performed using the steepest descent and conjugate gradient algorithms, converging when the maximum force on any atom fell below 100 kJ /mol nm. All simulations employed real-space Particle-Mesh-Ewald (PME) electrostatics and atom-based summation for van der Waals interactions with a cut-off distance of 1.2 nm. The time step for all simulations was set at 1.0 fs. Preliminary system equilibration was executed under a constant number of particles, pressure, and temperature (NPT) ensemble for 1000 ps using a velocity-rescale Berendsen thermostat, a temperature coupling constant set to 0.1 ps, and a semi-isotropic Berendsen barostat, with a pressure-coupling constant of 1 ps. The Berendsen thermostat and barostat facilitated rapid system equilibration, confirmed by convergence in the d-spacing and potential energy. Semi-isotropic pressure coupling allowed for effective computational decoupling of volume fluctuations in the $z$-direction and the $xy$-surface. For isolated monomer hydrocarbons, we set the temperature of the system based on geological temperature and pressure conditions to be constant at 323 K, with a constant pressure of 10 MPa. Notably, periodic boundary conditions were employed to achieve a sufficiently large system, and time-averaging was utilized to obtain macroscopic thermodynamic parameters. After preliminary equilibration, extended equilibrium simulations were conducted in the NPT ensemble for 20 ns using a velocity-rescale Berendsen thermostat and a semi-isotropic Parrinello–Rahman barostat, with both temperature and pressure coupling constants set to 1 ps.

The multi-component simulation system was conducted according to the thermodynamic conditions of the geological environment of the Songliao Basin in China, with the system temperature set at a constant 348 K and pressure maintained at 20 MPa. Initially, simulations lasting 400 ns were conducted under the NPT ensemble. However, multiple simulations reflected varying results, primarily due to the tendency of some hydrocarbon molecules to become trapped in local potential minima within the complex component simulation system. To better exhibit the competitive adsorption characteristics among mixed hydrocarbon components and prevent local entrapment, a further annealing simulation lasting 200 ns was carried out for the mixed component systems. The annealing protocol is set to exhibit periodic fluctuations in temperature, which it maintains at for a duration of 30 ns. It rises from 348 K to 473 K in three steps (348 K – 373 K – 423 K – 473 K), then cools down to 373 K (473 K – 423K – 373K), and finally returns to 348 K and holds for another 50 ns. Throughout the annealing process, the pressure of the system was maintained at 20 MPa, and the method for assessing the convergence of the complex mixed component system to an equilibrated state is further detailed in Section 2.4.1. After confirming the system reached equilibration, production simulations were carried out, as detailed in the pure-component systems above.

## 2.4 Analysis of molecular simulations

### 2.4.1 Assessment of system convergence

Insights obtained by MD have been established not only as a tool to interpret experimental results but also as a predictive instrument offering design insights for novel experiments [74,75]. However, the accuracy of MD simulation results and their representation of macroscopic states largely hinge on the convergence of the simulated system. Such convergence typically does not imply that the entire simulation system achieves a thermodynamic equilibrium state. Instead, through MD simulations, scholars can obtain the ensemble average of experimentally observable quantities by averaging over the dynamics of the model over time, thus reflecting the macroscopic thermodynamic state of the simulation system. For instance, the organic hydrocarbon components in oil reservoirs evolve slowly over geological timescales and both experiments and well equilibrated simulations can only represent the current state of organic matter evolution within the reservoir. According to the ergodic theorem, an infinitely long simulation would, in an ideal case, represent macroscopic phenomena observed experimentally. However, even with steady advancements in computational hardware and algorithmic efficiency expanding the temporal scope of simulations, we can never assume a system to be ergodic [76–78].

Past works frequently employed the convergence of the root-mean-square displacement (RMSD) relative to the initial configuration to judge system equilibration [79]. However, for systems containing a significant number of solvent molecules, its accuracy can be challenged. It is noteworthy that in simulations, the maximum RMSD can only be achieved when every molecule has moved a distance corresponding to the simulation box size, thereby fully exploring all available phase space. Moreover, RMSD is a global attribute, but in interface simulations, scholars often focus more on local interactions, such as near interface regions, defects, or specific surface sites. Given the constrained dynamics in these areas, they tend to evolve at a slower rate, making the use of RMSD for system equilibration assessment less accurate. Instead, in this article, we employed a method using the DynDen software to dynamically characterize and evaluate the convergence of hydrocarbon systems, encompassing both isolated monomers and multi-component mixtures, at the kaolinite basal surfaces [80]. Fig. S3 presents an example of the convergence assessment for isolated organic hydrocarbon systems using the DynDen software. The basis of this method is the assessment of linear density profile evolution (i.e., the mass density of individual organic components as a function of distance, orthogonal to the interfacial region). The system's convergence is then correlated with all-time frames' pairwise correlation coefficients [81]. RMSD may be insufficient to measure simulated equilibration due to the system's complexity, but this method effectively circumvents such obstacles. In this study, we evaluated the convergence of complex mixed systems by analysing the evolution of the density of different hydrocarbon components over simulation time, as illustrated in SI Fig. S4 and S5, based on the DynDen software. The reader is directed to Degiacomi *et al.* for a comprehensive explanation of this methodology.

#### 2.4.2 Spatial Density Distribution

The MD-obtained molecular trajectories contain the microscopic information of all molecules in the statistical ensemble, which may be translated into the appropriate macroscopic properties by means of statistical thermodynamics. The density distribution is a visual representation of the substance's essential properties and a macroscale representation of its conformational distribution at the microscopic level. In larger simulation systems, a statistical mechanics-based approach was used, where the system was segmented spatially into multiple bins, with the bin quantity contingent on the molecular count within the fluid. Regarding the confined fluid in the interlayer space between parallel kaolinite surfaces in this study, an indicator function $H_n$ was delineated to ascertain if a molecule is part of a designated bin [82]:

$$\begin{cases} H_n(Z_{i,j}) = 1 & (\mathrm{n}-1)\Delta Z < Z_i < n\Delta Z \\ H_n(Z_{i,j}) = 0 & Otherwise \end{cases} , \quad (1)$$

where the subscript $i$ and $j$ indicates the $i$ or $j$-th time step; $\Delta Z$ is the width of each bin. This function allows for the determination of the number of molecules located in the $n$-th bin. Then, for the $n$-th bin, the mean number density, $\rho_{num}$, and mass density, $\rho_{mass}$, distribution from $J_N$ to $J_M$ time steps of the hydrocarbons can be expressed as follows:

$$\rho_{num} = \frac{1}{A \cdot \Delta Z \cdot (J_M - J_N + 1)} \sum_{J=J_N}^{J_M} \sum_{i=1}^{N} H_n(Z_{i,j}) , \quad (2)$$

$$\rho_{mass} = \frac{10^{24}}{N_A} \frac{1}{A \cdot \Delta Z \cdot (J_M - J_N + 1)} \sum_{J=J_N}^{J_M} \sum_{i=1}^{N} H_n(Z_{i,j}) \cdot W_i , \quad (3)$$

where $Z_i$ denotes the coordinates of the midpoint for the $n$-th bin, while $A$ represents the area expressed in terms of $L_x \times L_y$. $J_N$ and $J_M$ correspond to the initial and final time steps for average parameter values respectively. $\rho_{mass}$ indicates the local mass density, in kg/m$^3$; $W_i$ represents the molecular weight, expressed in atomic mass units (u); $N_A$ stands for Avogadro's constant. On the basis of the hydrocarbon density, the average density of the monolayer adsorption phase for different adsorption layers can be estimated as

$$\rho_{ali} = \frac{m_{ali}}{V_{ali}} = \frac{\int_{L_i}^{L_{ii}} A_m \cdot \rho dL}{(L_{ii} - L_i) \cdot A_m} = \frac{\int_{L_i}^{L_{ii}} \rho dL}{L_{ii} - L_i} , \quad (4)$$

where $\rho_{ali}$ is the average density of the monolayer adsorption phase, in kg/m$^3$; $\rho$ is the number density or mass density distribution of the hydrocarbons; $l_n$ is the number of adsorption layers, $i$=1, 2, ... , n; $A_m$ is the pore specific surface area, in nm$^2$. Then, the average density in of the hydrocarbon's adsorbed and free phases is:

$$\rho_a = \frac{\sum_{i=1}^{n} \rho_{ali}}{n} , \quad (5)$$

$$\rho_f = \frac{m_f}{V_f} = \frac{\int_{L_1}^{L_2} A_m \cdot \rho dL}{(L_2 - L_1) \cdot A_m} = \frac{\int_{L_1}^{L_2} \rho dL}{L_2 - L_1} , \quad (6)$$

where $\rho_a$ is the average density of the adsorbed phase, in kg/m$^3$; $\rho_f$ is the average density of the free phase, in

kg/m$^3$.

### 2.4.3 Adsorption parameters of hydrocarbons

The determination of adsorbed layers on clay basal surfaces is intrinsically linked to adsorption parameters calculated based on the spatial density distribution of hydrocarbon fluids. Consequently, this study introduced a discriminant function, $\sigma_{df}$, specifically devised for determining the number of adsorption layers:

$$\sigma_{df} = \frac{\rho_{ali}-\rho_f}{\rho_f} \times 100\% \ . \tag{7}$$

In this study, a $L_i$ layer is defined as an adsorption layer when the fluid density within $L_i$ experiences a variation exceeding 2% relative to the fluid density of the bulk phase, that is, the discriminant function $\sigma_{df}$> 2%. It should be noted that the $\rho_{ali}$ and $\rho_f$ in Eq. (7) are determined based on calculations from Eqs. (4) and (6), utilizing data from the molecular number density distribution.

Based on the determination of the number of adsorption layers, the total thickness, $H$, of the adsorbed hydrocarbons can be expressed by the following equation:

$$H = \sum_{i=1}^{n} t_{ai} \ , \tag{8}$$

where $t_{ai}$ is the monolayer adsorption thickness, in nm; $n$ is the total number of adsorption layers, $i$=1, 2, ..., $n$. Previous works have demonstrated that the adsorption strength of inorganic minerals, such as clay, toward organic hydrocarbon fluids and gases like methane is primarily related to the surface area of the solid wall surface [83,84]. Therefore, the adsorptive capacity per unit surface area can indicate the intensity of interactions between solid wall surfaces and interior fluids. This work introduces the adsorptive capacity per unit area $C_a$ to characterize the organic hydrocarbon adsorption capability of the basal surfaces of kaolinite:

$$C_a = \frac{m_a}{A_m} = \frac{\int_{L_i}^{L_{ii}} A_m \cdot \rho_{mass} dL}{A_m} \ . \tag{9}$$

Where $A_m$ is the area of the solid wall surface, in m$^2$; $m_a$ is the mass of adsorbed molecules, in mg; $C_a$ is the amount of adsorption per unit area, mg/m$^2$.

### 2.4.4 The diffusion coefficients

Mean Squared Displacement (MSD) defines the particle displacement patterns within a system over time in dynamic simulations. Specifically, it assists in discerning whether particles are diffusing freely, transported, or restricted. The left side of Eq. (10) represents the MSD of hydrocarbons within the system, sampled every 1000 steps in the NPT ensemble simulation for hydrocarbon components. The random walk of particles in a uniform fluid is widely considered the standard method for calculating diffusion coefficients [85], as elucidated by the Einstein relation on the right side of Eq. (10):

$$\frac{1}{O}\frac{1}{O_t}\sum_{i=1}^{O}\sum_{O_t}\langle|r_i(t)-r_i(0)|^2\rangle = 2dDt \quad , \tag{10}$$

where $O$ represents the number of atoms in the hydrocarbons, and $O_t$ denotes the total number of time steps. $r_i(t)$ is the centroid position of the $i$-th atom at time $t$; $d$ signifies the spatial dimension considered for particle diffusion. For this study, particle diffusion in three-dimensional space is considered, hence $d = 3$.

### 2.4.5 Hydrocarbons – clay interaction energy

In the process of conducting MD simulations, we define energy groups to analyze the non-bonded interaction energies between various hydrocarbons and kaolinite surfaces through a rerun of the trajectories. Typically, only non-bonded interactions including van der Waals forces (represented by the 12-6 Lennard-Jones potential) and long-range Coulomb interactions are considered, as depicted by Eq. (11). Initial simulations are conducted utilising the PME method for periodic dynamical investigations to ensure the accuracy of interaction energy computations. Subsequently, the cutoff approach is utilised to calculate electrostatic interactions. The designation of energy groups is followed by a rerun of the simulation trajectories.

$$E_{nonbond} = \sum_i E_i^{Coul} + \sum_{j>i} E_{i,j}^{Coul} + \sum_i E_i^{VDW} + \sum_{j>i} E_{j>i}^{VDW}, \tag{11}$$

Where $i$ and $j>i$ represent non-bonded intramolecular and intermolecular interactions, respectively. This study focuses solely on the energies of intermolecular interactions, disregarding the intermolecular interactions energy component. In particular, $E_{i,j}^{Coul}$ and $E_{i,j}^{VDW}$ represent the long-range intermolecular Coulomb and van der Waals interaction terms, respectively.

Typically, the free energy change of a system can be divided into two parts: the change in free energy under ideal conditions and the excess free energy change caused by intermolecular interactions. Both parts can be calculated by applying formalized density functional theory, rather than relying solely on experimental measurements [86,87]. However, for adsorption systems on kaolinite surfaces, using density functional theory to calculate the interfacial adsorption free energy is challenging because the kaolinite in this study is modeled as a real structure composed of silicon-oxygen tetrahedra and aluminum-oxygen octahedra. Since the MD simulations in this study have sufficiently sampled the system space, the local density of hydrocarbon fluid molecules will be directly related to the probability of a molecule occupying that system space [88]. A good approximation of the system partition function can be obtained by normalizing the molecular number density. The change in free energy of the normalized organic hydrocarbon molecular probability density is calculated based on Boltzmann inversion using the following equation [89,90]:

$$U(z) = -k_B T ln\left(\frac{p(z)}{p_0}\right) \quad , \tag{12}$$

where $k_B$ is the Boltzmann constant, $T$ is the temperature of the simulated system, $U(z)$ represents the free energy change of the system at position z, $\rho(z)$ and $\rho_0$ are the fluid densities at the kaolinite surface position z and in the intermediate bulk phase region, respectively. The free energy change calculated using Eq. (12) can offer a quantitative description of the energy stability of the adsorption configuration and an assessment of barrier penetration between adsorption layers in systems containing a large number of molecules.

#### 2.4.6 Alignment of organic hydrocarbon molecules as a function of distance from the surface

To elucidate the interactions between organic hydrocarbon molecules and the basal surface of kaolinite, we characterized the molecular orientation in relation to the surface normal, contingent on the structure of the hydrocarbon molecules, employing vectors or predefined surfaces as descriptors. For straight-chain alkanes, vectors are designated based on the orientation of the molecule or the bond involving the most electronegative atom, as illustrated in the shaded regions of the molecular configurations in Fig. 5(a) and (b). In the case of aromatic hydrocarbons or asphaltenes, the surface is determined by their aromatic ring structures, thus more accurately representing the orientation characteristics of these molecules on the kaolinite surfaces [91].

Our orientation descriptors require the atomic coordinates of all atoms constituting the hydrocarbon molecules. These coordinates are extracted from the molecular trajectory file and archived utilizing a Tcl script in VMD 1.9.4 [73]. Thereafter, these coordinates undergo processing via an in-house Python script to deduce orientation details for each hydrocarbon molecule at individual time frames. The extraction technique incorporates vectors and surface normals in spherical coordinates, coupled with the mean *z*-coordinate position of the hydrocarbon molecules. Leveraging this data, we constructed 2D histograms that exhibit the distribution of each orientation descriptor (the elevation and azimuth angles of vectors and surfaces) in correlation with their *z*-axis positioning (Fig. 5).

## 3. Results and Discussions

### 3.1 Adsorption characteristics of pure hydrocarbons on kaolinite basal surfaces

#### 3.1.1 Adsorption capacity and molecular alignments of hydrocarbons

In this study, the adsorption configurations of different organic hydrocarbon components in the vicinity of varied basal surfaces of kaolinite were simulated. As depicted in Fig. 3, the fluid density distribution was determined using Eq. (3) as a function of the distance *z* from the clay basal surface in the direction perpendicular to the surface. The density curves for the basal HY and SI surfaces of kaolinite are depicted on the left and right sides of the graph, respectively. The density of different hydrocarbon monomers near the kaolinite basal surface fluctuates to varying degrees, indicating that hydrocarbons are adsorbed near the surface due to interaction with the surface, which is consistent with previous studies [27,29]. Moreover, the density distribution of different components also shows significant differences in the adsorption characteristics and distribution patterns near the two basal surfaces of the kaolinite. Fig. 3(a) shows the mass density distribution of saturated alkanes with different carbon numbers from n-hexane to n-eicosane. The figure illustrates that an increase in the carbon number corresponds to a progressive increase in the density peak of various adsorption layers, an increase in the amount of adsorption, and a tendency for the bulk phase density to increase in tandem with the carbon number. Additionally, it was noted that the adsorption density of saturated hydrocarbons is greater in the vicinity of the SI basal surface of kaolinite compared to the HY basal surface. For instance, the first adsorption layer density peak of n-eicosane on the HY surface is lower than that on the SI surface (i.e., 2.16 g/cm$^3$ vs 2.43 g/cm$^3$, respectively, highlighted by a horizontal blue dashed line on Fig. 3a). This demonstrates that adsorbed nonpolar organic matter on the SI basal surface of kaolinite is more readily possible, supporting the hydrophobic character of the SI surface. The parameters of hydrocarbon monomer adsorption on the basal surface of kaolinite, listed in Table 2, show that the bulk phase density of saturated hydrocarbons with different carbon numbers is generally consistent with experimentally measured densities, validating the equilibrium of the simulation system and the accuracy of the simulation results.

The dark blue dashed line in Fig. 3(a) illustrates the distribution of squalane densities. Although its carbon number reaches 30, the mean peak value of the density of its first adsorption layer is comparatively lower compared to n-eicosane (1.59 g/cm$^3$ vs 2.43 g/cm$^3$). This indicates that the presence of branches in saturated hydrocarbons disrupts the adsorption layer characteristics distributed parallel to the basal surface, making it more challenging for them to overcome spatial effects and adsorb near the clay basal surface compared to that of straight-chain alkanes. The bulk phase density of squalane, however, is higher compared to n-eicosane, with an average value of 0.84 g/cm$^3$ closely matching the experimental value of 0.81 g/cm$^3$, which further validates the simulation soundness.

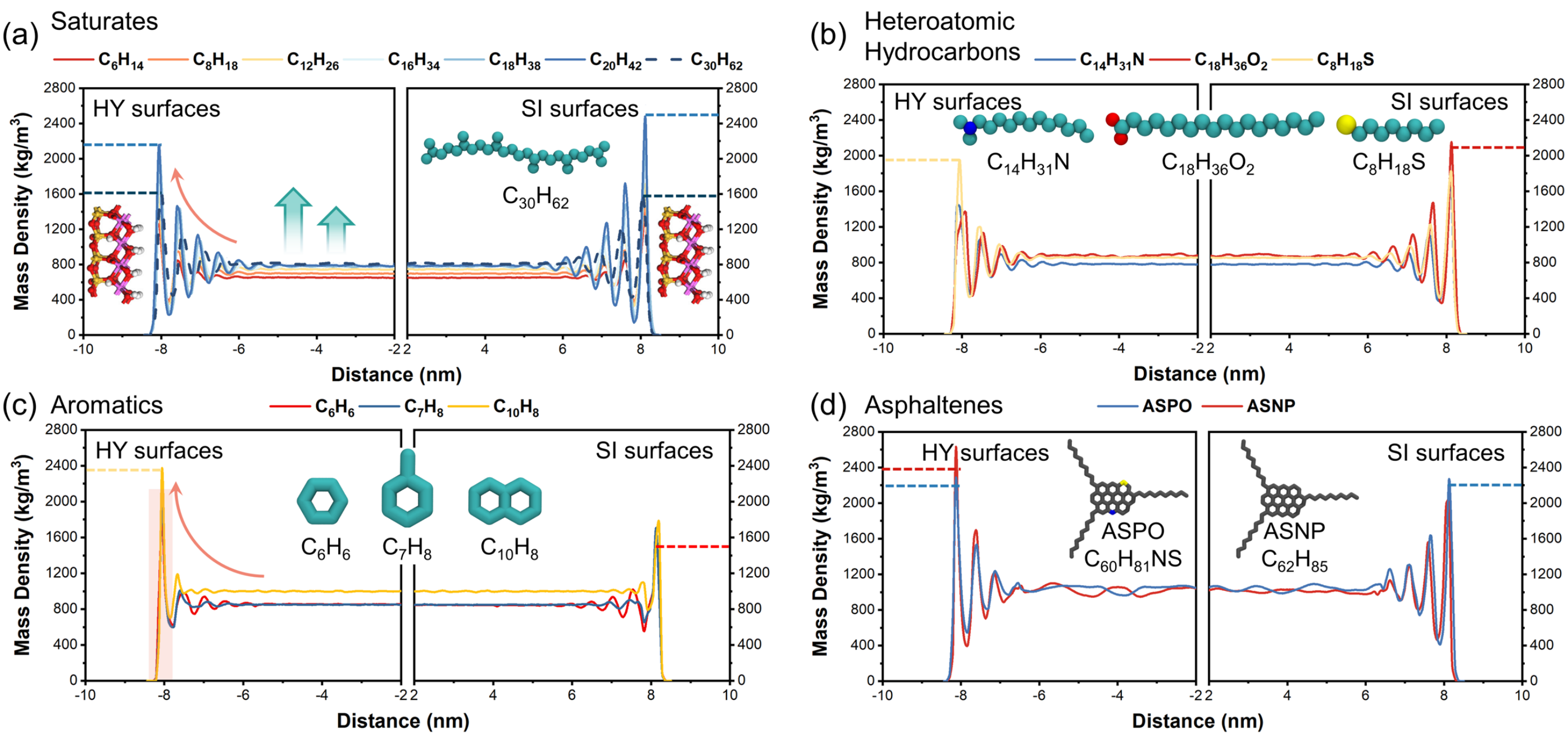

***Fig.3.*** *Density distributions of isolated hydrocarbons between two basal surfaces of kaolinite. Mass density distribution of (a) saturated hydrocarbons with different carbon chain lengths along the z-axis, where the inset shows schematic molecular models of kaolinite (representing two different clay basal surfaces) and squalane; (b) three heteroatomic hydrocarbons and their schematic molecular models; (c) three aromatic hydrocarbons alongside their corresponding molecular models; (d) two types of asphaltenes and their models.*

The study compared the density distributions of three different heteroatomic hydrocarbons (Fig. 3(b)), revealing that stearic acid, an oxygen-containing compound, exhibits the highest adsorption density. Its first density peak in the adsorption layer on the siloxane surface of kaolinite is higher than the peak density on the hydroxylated surface (i.e., 2.24 g/cm$^3$ vs 1.54 g/cm$^3$, respectively). While polar organic molecules are more easily adsorbed onto the hydroxylated basal surface of kaolinite, compounds such as stearic acid, which have an amphiphilic molecular structure and an oxygen-containing functional group that is polar at the head and nonpolar at the straight-chain alkane tail, exhibit an adsorption configuration on the hydroxylated basal surface that is distinguished by the simultaneous presence of perpendicular and parallel orientations relative to the basal surface. This molecular arrangement, compared to the parallel adsorption interactions on the siloxane basal surface, results in weaker interactions, allowing hydrocarbon molecules to escape the control of the basal surface and enter the bulk phase more easily. Consequently, under the influence of spatial effects, the average adsorption amount is low. On the contrary, the adsorption densities of nitrogen and sulphur-containing compounds are progressively lower than those of oxygen-containing compounds. Among them, the sulfur-containing compound 1-octanethiol has the lowest adsorption density peak on the siloxane basal surface, at 1.78 g/cm$^3$. This is attributed to sulfur's electronegativity being very similar to that of carbon and hydrogen, contributing less to polarity. Therefore, its adsorption characteristics on both basal surfaces of kaolinite are essentially consistent, indicating that its adsorption configuration is primarily parallel.

From Fig. 3(c), it can be seen that the density distribution characteristics of aromatic hydrocarbon compounds are relatively consistent, with smaller fluctuations in their density curves near the basal surface, indicating fewer adsorption layers compared to other hydrocarbon components. Based on the density curves, it is evident that aromatic hydrocarbon components with aromatic rings are more likely to adsorb on the polar hydroxyl surface (left side of Fig. 3(c)). For example, the density peak of naphthalene on the hydroxylated surface is 2.84 g/cm$^3$, which is significantly higher than the peak density of 1.67 g/cm$^3$ at the siloxane surface. Although hydrocarbons such as benzene and naphthalene are typically considered non-polar, their π-electron clouds generate transient dipole moments that promote their adsorption on the hydroxylated surfaces of kaolinite. While classical force fields do not account for explicit electrons, the charge distribution on the atoms allows to account for this. Moreover, by comparing the density curves of methyl-substituted toluene, we find that the adsorption density peaks of this compound near the basal surface of the polar clay are higher than those of benzene, revealing the complexity of surface adsorption. The height of the density peaks reflects the degree of molecular aggregation or local adsorption strength in specific surface areas. Methyl substituents may play a positive role in local interactions with polar surfaces. However, due to the larger molecular volume and increased molecular weight of methyl-substituted aromatic hydrocarbons, they exhibit higher local density values. Still, the overall adsorption amount on the clay basal surface is lower than that of aromatic hydrocarbons without methyl substitution (Table 2) due to spatial effects.

For more complex hydrocarbons with multiple aromatic rings, such as asphaltenes, their adsorption characteristics on the clay surfaces differ significantly from those of small-molecule aromatics. As shown in Fig. 3(d), both types of asphaltene exhibit multilayer adsorption characteristics on the kaolinite basal surfaces, and there is no significant difference in the adsorption characteristics of the two asphaltenes on the two types of the surface. The density curves show that heteroatom-containing asphaltene molecules (ASPO) have a slightly higher adsorption density peak at the first adsorption layer on the hydroxylated surface (2.61 g/cm$^3$) compared to non-heteroatom asphaltenes (ASNP) with a density peak of 2.26 g/cm$^3$. This is because asphaltene molecules containing sulfur and nitrogen atoms have local electronegativity differences on their aromatic rings, leading to stronger interactions with the polar surfaces of clay minerals through dipole-dipole interactions, like seen for the heteroatomic hydrocarbons (Fig, 3(b)). However, since heteroatoms in asphaltenes are known to be relatively uniformly distributed, there are no strongly polar molecules. Therefore, for asphaltenes composed primarily of nonpolar hydrocarbons, the main interaction remains London dispersion forces, which are a type of van der Waals interaction [92].

To elucidate the adsorption behaviour and molecular configuration distribution characteristics of various hydrocarbons on different basal surfaces of kaolinite, in this work, we employed time-averaged calculations in a state of equilibrium to ascertain the number density distribution in the *x*- and *y*-axes of the primary adsorption layer on both basal surfaces of the clay. The renderings of the molecular configurations of the first adsorbed layer are shown in columns (a) and (c) of Fig. 4), while the bidimensional density distributions are shown in columns (b) and (d) of Fig. 4. The left-hand columns (Fig. 4(a) and (b)) illustrate the molecular configuration traits of the first adsorption layer on the hydroxyl basal surface of kaolinite, whereas right-hand columns (Fig. 4(c) and (d)) present the distribution on the siloxane basal surface. The molecular configuration characteristics of the first adsorption layer of saturated hydrocarbons, exemplified by *n*-eicosane, reveal that the adsorbed hydrocarbons are uniformly aligned parallel to the clay's basal surface, exhibiting a predominant molecular orientation along the hexagonal symmetry axes of the siloxane lattice. This pattern is in complete agreement with previous studies [28,37,93]. Furthermore, prior research that uses empirical techniques such as transmission electron microscopy and atomic force microscopy has also observed the parallel arrangement of organic hydrocarbon adsorption layers [94–96], thereby corroborating that the simulation results of this study accurately represent the distribution patterns of hydrocarbons on clay basal surfaces.

With respect to the distribution of the first adsorption layer of amphiphilic molecules like stearic acid, it is apparent that the molecular arrangement on the siloxane basal surface of kaolinite is more structured and denser compared to that on the hydroxyl basal surface. The stearic acid molecules on the siloxane basal surface predominantly align parallel to this surface (heteroatom hydrocarbon distribution characteristics depicted in Fig. 4(c) and (d)), with molecular orientations corresponding to the hexagonal lattice axes of the siloxane, akin to the arrangement observed with *n*-octadecane. Fig. 4(a) and (b) demonstrate that the stearic acid molecules in the initial adsorption layer on the hydroxylated basal surface display their oxygen-containing functional groups, with some molecules interacting with the basal surface solely through their oxygen-containing functional group heads. This occurs because the polar oxygen-containing functional heads of stearic acid form hydrogen bond interactions with the hydroxylated basal surface, resulting in an orientation perpendicular to the surface.

Fig. 4 illustrates the molecular arrangement attributes of the initial adsorption layer of aromatic hydrocarbons and asphaltenes on both basal surfaces of kaolinite. Analysis of Fig. 4(a) and (b) reveals that on the hydroxylated basal surface of kaolinite, molecules of aromatic hydrocarbons and asphaltenes predominantly align parallel to the basal surface, accounting for the elevated adsorption density peaks of aromatics and asphaltenes on this surface. The polarity of kaolinite's hydroxylated basal surface facilitates weak electrostatic interactions with the π-electron clouds of the aromatic rings in aromatic hydrocarbons [97], thus promoting the parallel orientation of the aromatic ring portion to the hydroxylated surface to optimize this interaction. Furthermore, when subjected to systemic pressure, the aromatic ring-containing hydrocarbon molecules are combined with the partial repulsion of electron clouds caused by the hydroxyl groups on the surface to overcome spatial constraints. This results in the formation of a stable parallel arrangement structure through π-π stacking interactions [98]. Conversely, the siloxane basal surface of kaolinite, characterized by its reduced surface electronegativity and absence of pronounced polar centers, is less effective in engaging with the π-electron clouds of aromatic hydrocarbon molecules. As a result, the molecular arrangement under van der Waals forces exhibits greater randomness, and the pattern of parallel alignment of aromatic ring portions to the basal surface is less prevalent (as depicted in Fig. 4(c) and (d)).

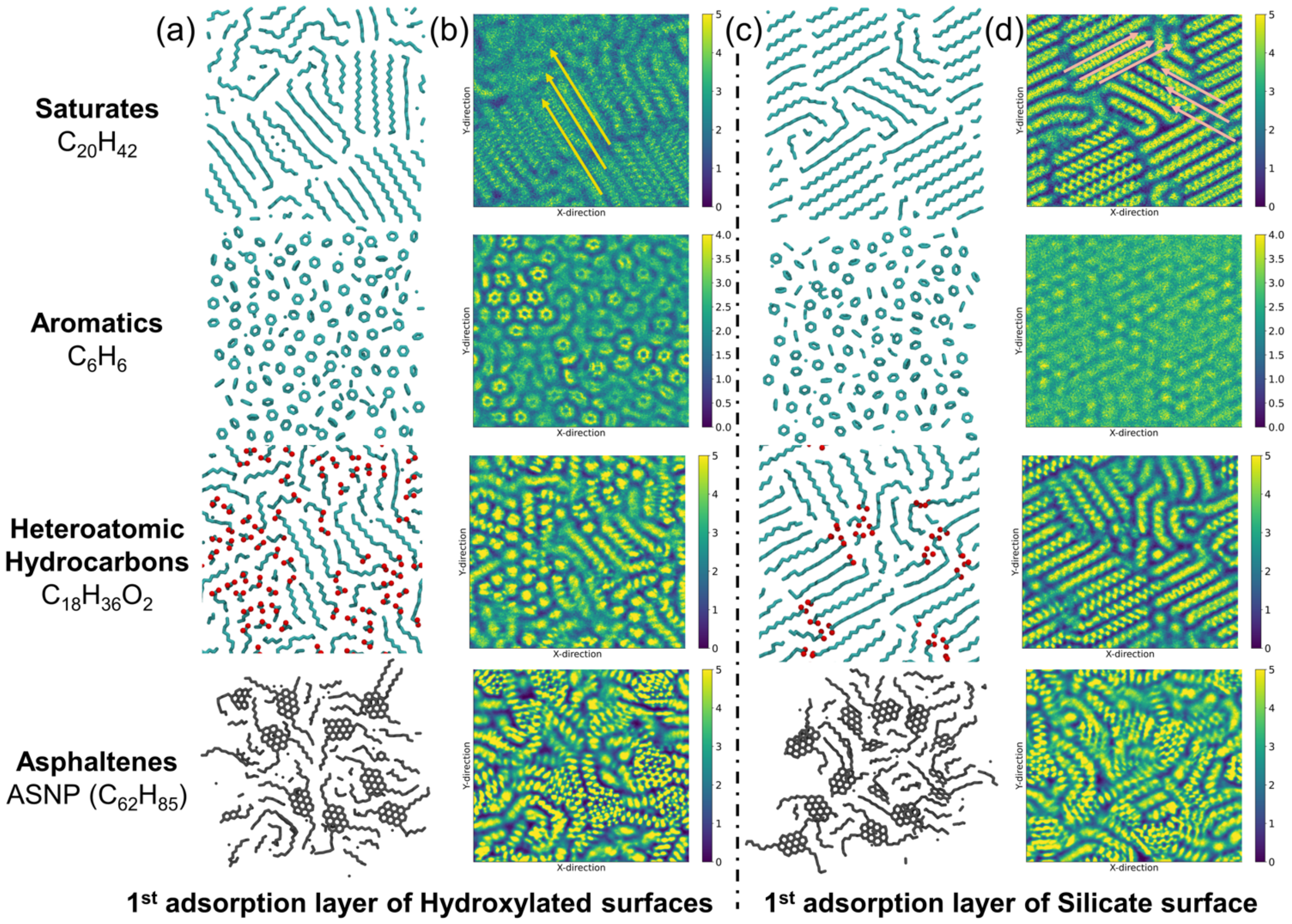

***Fig. 4.*** *Molecular configuration and 2D density distribution of the first adsorbed layer of isolated hydrocarbons on two basal surfaces of kaolinite (a) molecular configuration distribution of four hydrocarbon components in the 1*$^{st}$ *adsorption layer on the kaolinite HY surface; (b) density distribution of the1*$^{st}$ *adsorption layer on the HY surface, where high-density areas are represented in yellow, low-density areas in blue, and the alignment of saturated hydrocarbons is indicated by yellow arrow lines; (c)*

*configuration distribution of hydrocarbons on the SI surface, with the alignment of saturated hydrocarbons indicated by orange arrow lines; (d) density distribution of four components on the SI surface. On (a) and (c) all hydrogen atoms are hidden for clarity, red represents oxygen atoms, and remaining atoms as carbon.*

Through the examination of the orientation and distance relationships of hydrocarbon molecules, this study investigated the alignment of various hydrocarbon components on the basal surfaces of kaolinite. Fig. 5(a) elucidates the distribution traits of elevation and azimuth angles of the straight-chain saturated hydrocarbon *n*-eicosane on the surfaces of kaolinite. The findings reveal that saturated hydrocarbons predominantly align parallel to the surface, facilitating multilayer adsorption characteristics. When the azimuth angle distribution is analysed, it is evident that the molecules exhibit a relatively uniform orientation, primarily aligning along the hexagonal symmetry axis of the siloxane lattice. However, owing to the distinct properties of kaolinite's basal surfaces, there is a variation in the azimuth angles of the alkanes. Fig. 5(b) shows the distribution of elevation and azimuth angles of the head of the oxygen-containing functional group of stearic acid. The distribution of elevation angles is concentrated primarily around 80 degrees and –40 degrees, substantiating that the polar oxygen-containing functional group head of stearic acid establishes hydrogen bonds with the hydroxylated basal surface, resulting in the molecules orienting perpendicular to the surface. The azimuth angle distribution demonstrates that stearic acid is exposed exclusively on the hydroxylated surface of kaolinite, displaying a random distribution, whereas on the siloxane surface, the molecules arrange parallel to the surface, rendering the orientation of the oxygen-containing functional groups nearly imperceptible (as depicted in Fig. 4(c)).

The elevation angle distribution of the benzene aromatic ring surface on kaolinite's basal surfaces indicates that benzene primarily forms adsorption layers parallel to both surfaces (Fig. 5 (c)). The azimuth angle distribution reveals that the aromatic hydrocarbons are randomly orientated on the surface. The elevation angle distribution of non-substituted asphaltenes, ASNP, suggests that they predominantly adsorb onto the surface with their polycyclic aromatic structures aligned parallel to the clay surface, exhibiting no significant disparity in molecular orientation between the two surfaces (Fig. 5 (d)). Therefore, the azimuth angle distribution of asphaltenes indicates a random arrangement in the clay.

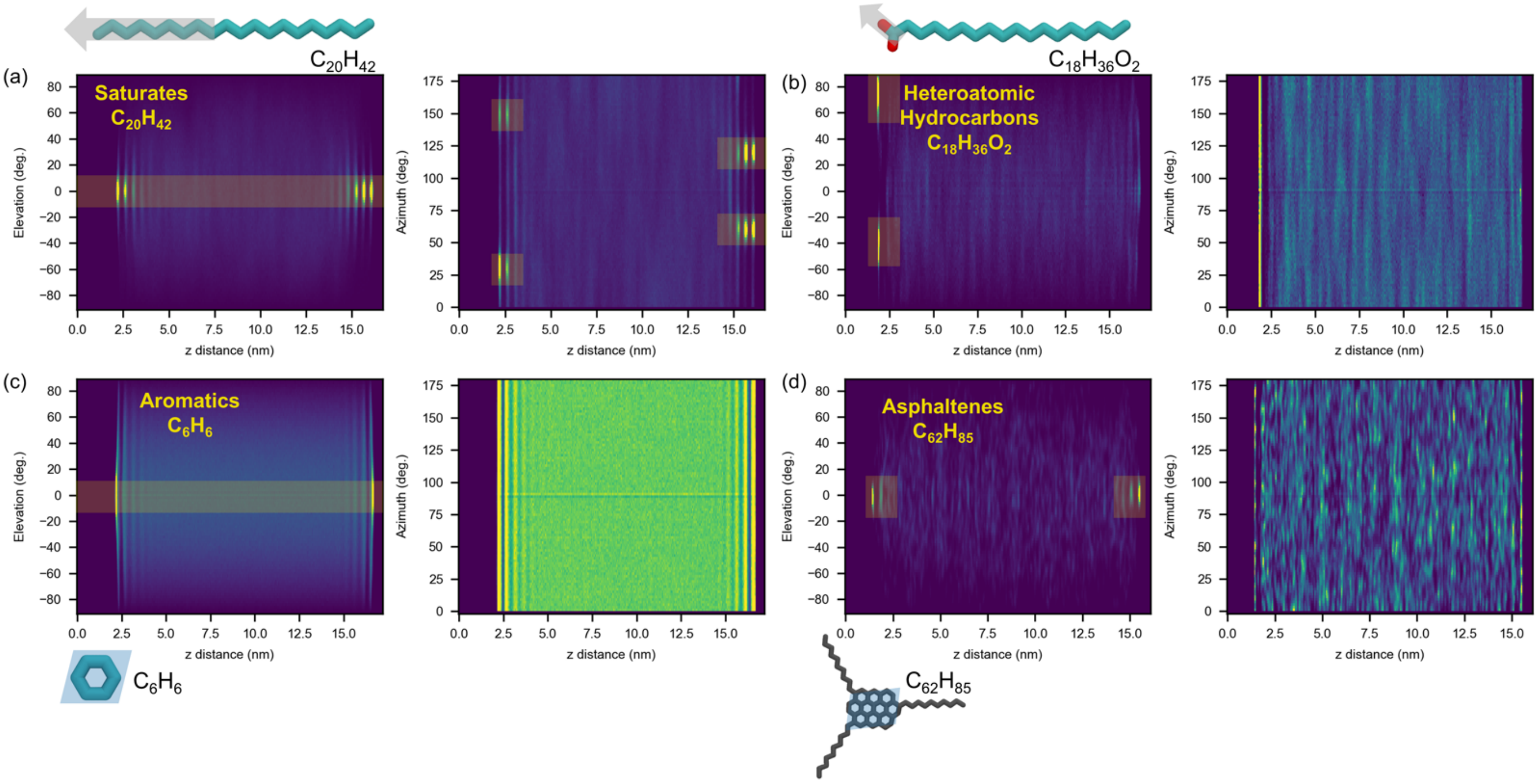

***Fig.5.*** *Statistical distribution of orientations of various hydrocarbon molecules on kaolinite basal surfaces (a) distribution characteristics of the elevation and azimuth angles of n-eicosane on the kaolinite HY and SI basal surfaces, where the arrows in the inset indicate the direction used to statistically analyse molecular orientations; (b) alignment distribution of octadecanoic acid on the two basal surfaces, with arrows in the inset showing the molecular orientation along the oxygen-containing functional group; (c) alignment distribution of benzene on the two basal surfaces, where the blue diamond in the inset represent the molecular orientation based on the benzene ring plane; (d) alignment distribution of ASNP on the two basal surfaces, with diamonds in the inset indicating the molecular orientation based on the plane of the polycyclic aromatic hydrocarbon structure.*

### 3.1.2 Adsorption and diffusion characteristics of hydrocarbons

The adsorptive disparities and fluidic behaviours of various hydrocarbons on the basal surfaces of kaolinite can be quantitatively evaluated by examining the adsorption per unit area and the molecular diffusion coefficients. In this work, the adsorption quantities of isolated monomer hydrocarbons on kaolinite basal surfaces and their diffusion coefficients within the kaolinite pores were determined using Eq.s (9) and (10), with the findings illustrated in Fig. 6. For saturated hydrocarbons, we analysed the adsorption quantities of straight-chain alkanes with varying carbon chain lengths on kaolinite basal surfaces. Fig. 6(a) illustrates that from *n*-hexane to *n*-eicosane, the adsorption per unit area follows a linear increment trend, escalating from 1.31 mg/m² for n-hexane to 2.22 mg/m² for *n*-eicosane (Table 2). This aligns with the observations in Fig. 3(a), which indicate an elevation in adsorption density corresponding to an increase in carbon numbers and is consistent with prior works dedicated to alkane adsorption properties [27,99]. This means that an increase in the number of carbons increases molecular polarizability, thus amplifying dispersion interactions between the molecules and the basal clay surfaces, culminating in increased adsorption of saturated hydrocarbons on clay surfaces [100].

Fig. 6(b) examines the variations in the diffusion coefficients of alkanes with different carbon chain lengths within the pores of kaolinite, revealing a marked decline with increasing carbon numbers, decreasing from $3.09\times10^{-9}$ $m^2/s$ for *n*-hexane to $0.13\times10^{-9}$ $m^2/s$ for *n*-eicosane (Table 2). This suggests, unsurprisingly, that an increase in carbon chain length significantly decreases the diffusion rate of alkane molecules within clay pores, and long-chain saturated hydrocarbons progressively form a waxy crystalline state, substantially impeding their mobility. Moreover, this research investigated the fluid behaviour of squalane within clay pores, in contrast to saturated hydrocarbons with straight chains and carbon lengths that varied. Fig. 6(c) demonstrates that the adsorption of squalane per unit area on the clay surface is 1.69 $mg/m^2$, lower than that of *n*-eicosane and akin to *n*-dodecane at 1.62 $mg/m^2$. For squalane, with a carbon count of 30, its adsorption does not adhere to the linear increase pattern observed with carbon number, attributable to its unique molecular structure and shape. Straight-chain alkanes possess a linear long-chain molecular configuration, whereas squalane exhibits a complex branched framework. This branched architecture enlarges the molecular volume and introduces steric hindrances, diminishing effective adsorption interactions between squalane molecules and clay basal surfaces. Additionally, straight-chain saturated hydrocarbons form a compact, parallel crystalline structure on clay basal surfaces (as depicted in Fig. 4 and 5), facilitating multilayer adsorption of hydrocarbons on the clay surface. In contrast, due to squalane's intricate branched structure, its alignment on the clay surface is more sporadic, lessening the probability of multilayer adsorption. However, given that squalane molecules are larger in mass and the branched structure increases the volume, their mobility in the constrained clay pore space is minimal, leading to a reduced diffusion coefficient.

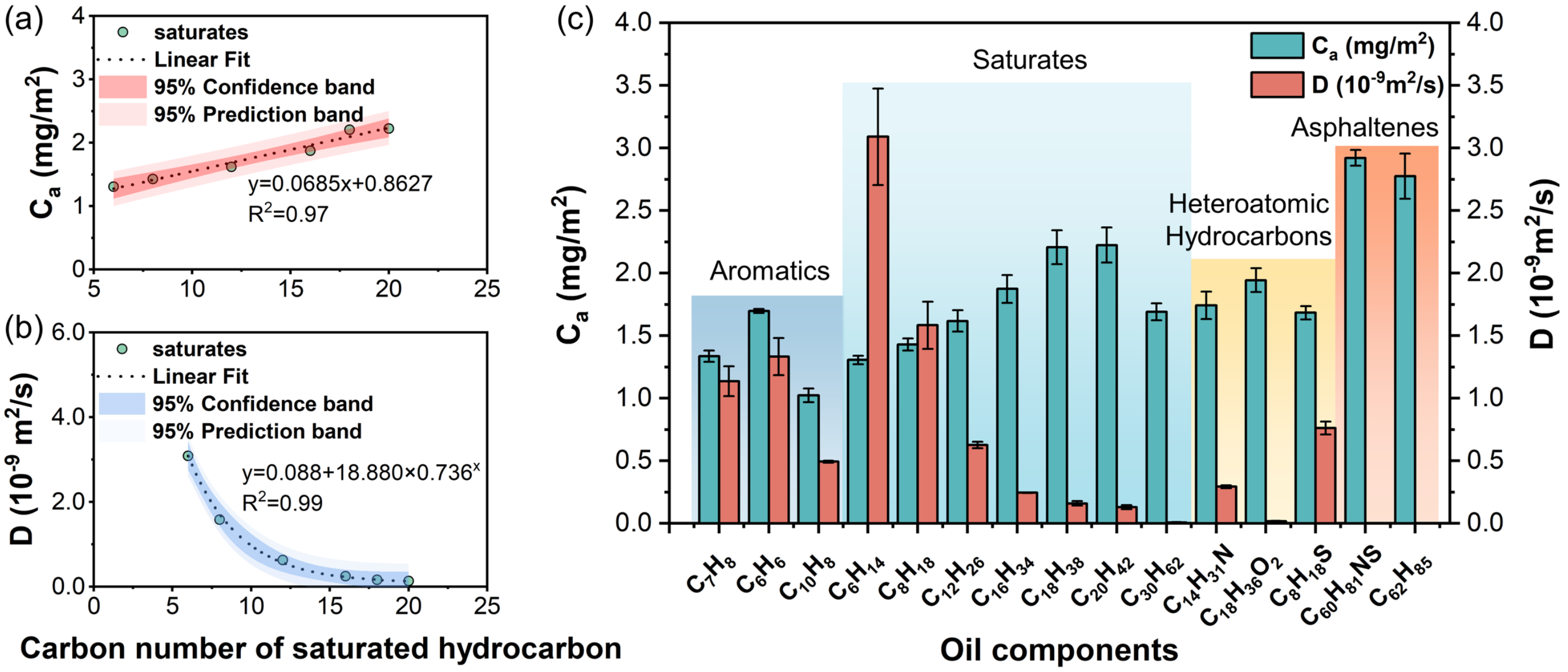

***Fig. 6.*** *Adsorption and diffusion characteristics of isolated hydrocarbons adsorbed on kaolinite basal surfaces (a) the variation of the adsorption amount per unit area of saturated hydrocarbons as a function of carbon chain length; (b) the diffusion coefficient of saturated hydrocarbons as a function of carbon chain length; (c) the distribution histograms of the adsorption amount per unit area and diffusion coefficient for different hydrocarbons, where the dark blue area represents aromatic hydrocarbons, the light blue area represents saturated hydrocarbons, the yellow area represents heteroatomic hydrocarbons, and the orange area represents asphaltenes.*

Fig. 6(c) presents the adsorption quantities of hydrocarbons with aromatic rings on kaolinite basal surfaces. In the context of the pure hydrocarbon system, the adsorption of benzene on kaolinite surfaces surpasses that of toluene, and the adsorption of toluene exceeds that of naphthalene. As depicted in the density distribution in Fig. 3(c), aromatic hydrocarbons on clay basal surfaces do not establish compact multi-layer adsorption as observed in straight-chain saturated hydrocarbons, leading to a marginally lower adsorption quantity than that of saturated hydrocarbons. Benzene, characterized by its mono-ring structure, occupies a reduced space and exhibits a more orderly molecular configuration, thereby facilitating its adsorption by the polar kaolinite basal surface. Due to the presence of the π-electron cloud, benzene can effectively form multi-layer adsorption, resulting in an augmented adsorption quantity on kaolinite basal surfaces. The methyl substituent in toluene enlarges its molecular volume and mass. Although this results in a higher local density, spatial constraints impede toluene's ability to develop multi-layer adsorption structures on the clay basal surface, culminating in a lower overall adsorption quantity compared to benzene. Furthermore, the increased molecular mass and volume from the methyl substituent in toluene contribute to a diminished diffusion coefficient relative to benzene. Naphthalene, as a bi-ring aromatic compound, possesses a greater molecular volume than benzene and toluene. Particularly in spatially constrained systems, the search of suitable adsorption sites on clay basal surfaces may be impeded by its larger size. Despite naphthalene's dual aromatic rings facilitating robust π-π stacking interactions, these interactions tend to cause naphthalene molecules to form smaller aggregates amongst themselves rather than with the clay basal surface, diminishing the probability of multi-layer adsorption. Consequently, its adsorption quantity is lower than that of benzene and toluene. The bi-ring structure of naphthalene exacerbates movement obstruction within clay pores, encountering more geometric constraints in constricted spaces, thus leading to lower diffusion rates.

Fig. 6(c) further illustrates that the two variants of asphaltenes exhibit the highest adsorption quantities on clay basal surfaces, substantially exceeding the other three components (2.77 mg/m$^2$ for ASNP and 2.92 mg/m$^2$ for ASPO). The adsorption quantity of heteroatom-inclusive asphaltenes is higher than that of non-heteroatom asphaltenes (Fig. 6(c) and Table 2). This phenomenon is attributed to asphaltene molecules containing sulfur and nitrogen atoms within their aromatic rings, generating localised electronegativity disparities and forming more potent interactions with kaolinite's polar surfaces via dipole-dipole interactions.

Based on the adsorption capacity per unit area (Fig. 6(c)), the rank of adsorption capacity among the three heteroatomic hydrocarbons and kaolinite is established as follows: stearic acid (1.94 mg/m$^2$), surpasses dimethyldodecylamine (1.74 mg/m$^2$), which in turn exceeds octanethiol (1.68 mg/m$^2$). Notwithstanding the tendency of kaolinite's polar basal surfaces to exhibit increased adsorption affinity towards stearic acid, which presents oxygen-containing functional groups, as shown in Fig. 4 and 5, the amphiphilic characteristics of stearic acid, nevertheless, engender its adsorption capacity to fall below that of higher-carbon-number straight-chain saturated hydrocarbons (1.94 mg/m$^2$ vs 2.21 mg/m$^2$). Moreover, the sequential adsorptive capacities of these compounds on the basal surfaces of kaolinite align with the relative electronegativities of the heteroatoms oxygen, nitrogen, and sulphur. From Fig. 6(c), it is seen that the diffusion coefficients of the heteroatomic hydrocarbons manifest an inverse relationship to their adsorptive capacities, wherein octanethiol demonstrates the most pronounced diffusion capability ($0.76\times10^{-9}$ m$^2$/s) compared to the other compounds, whereas the mobility of stearic acid within the confines of the pressure-controlled clay pores is virtually negligible (i.e., $0.02\times10^{-9}$ m$^2$/s).

### 3.1.3 Comparison of interaction energies across hydrocarbons

The affinity of various isolated monomer hydrocarbons for the basal surface of kaolinite is elucidated through thermodynamic energy parameters (Fig. 7). The non-bonded interaction energy between distinct hydrocarbons and the clay's basal surface was quantified utilizing Eq. (11). Displayed in Fig. 7(a) is a graphical representation comparing the intensity of interactions among diverse hydrocarbons and the clay basal surface, illustrating that the interaction energy between asphaltenes and the clay surface surpasses that associated with heteroatomic hydrocarbons, which in turn exceeds that observed with saturated hydrocarbons, and ultimately aromatic hydrocarbons. For nonpolar saturated hydrocarbons, encompassing both low and high-molecular weight alkanes, adsorption onto the kaolinite basal surface predominantly occurs via van der Waals forces, which these forces contribute to 99.7 % of the total molecular interaction energy. Such evidence reiterates the notion that nonpolar linear or branched alkanes elicit dispersion forces with the clay surface through van der Waals interactions, aligning with prior study [37,98]. Nonetheless, upon correlating the non-bonded interaction energy with the adsorption amount per unit area, one discerns that the adsorption capacity of saturated hydrocarbons is not solely contingent on the carbon atom count and molecular polarizability but is also influenced by molecular conformation and spatial effects (as illustrated in Fig. 7(b)). For instance, while the interaction energy between squalane and kaolinite exceeds that of *n*-eicosane (i.e., −147.9 kcal/mol), its adsorption amount is diminished in comparison to straight-chain alkanes with fewer number of carbon atoms, attributable to the branched conformation and spatial constraints of the molecule.

Aromatic hydrocarbons are adsorbed onto the clay basal surface through a combination of Coulombic and van der Waals forces (Fig. 7(a)). Within this domain, non-bonded interactions are substantially governed by van der Waals forces, constituting on average 74 % of the total forces. Nonetheless, the transient dipole moments engendered by the π-electron clouds inherent to aromatic hydrocarbons enhance their adsorptive affinity for the polar surface of kaolinite, contributing to a quarter of total interactions via Coulombic forces. These insights contribute to our understanding of the preferential adsorption orientation of aromatic hydrocarbons along the polar surface of kaolinite, as evidenced in Fig. 4 and 5. In the context of asphaltenes, an examination of their complex structures that incorporate polycyclic aromatic frameworks reveals that van der Waals forces continue to exert the greatest influence on their interactions with the clay basal surface, accounting for an average of 87 %. This is evident in the notably high adsorption amounts that asphaltenes exhibit in comparison to other components. The disparity in interaction energies between the two asphaltene variants is ascribed to marginally more pronounced Coulombic forces associated with the heteroatom-containing asphaltene ASPO when juxtaposed with ASNP, upon interaction with kaolinite's polar surface, as shown in Fig. 7(b).

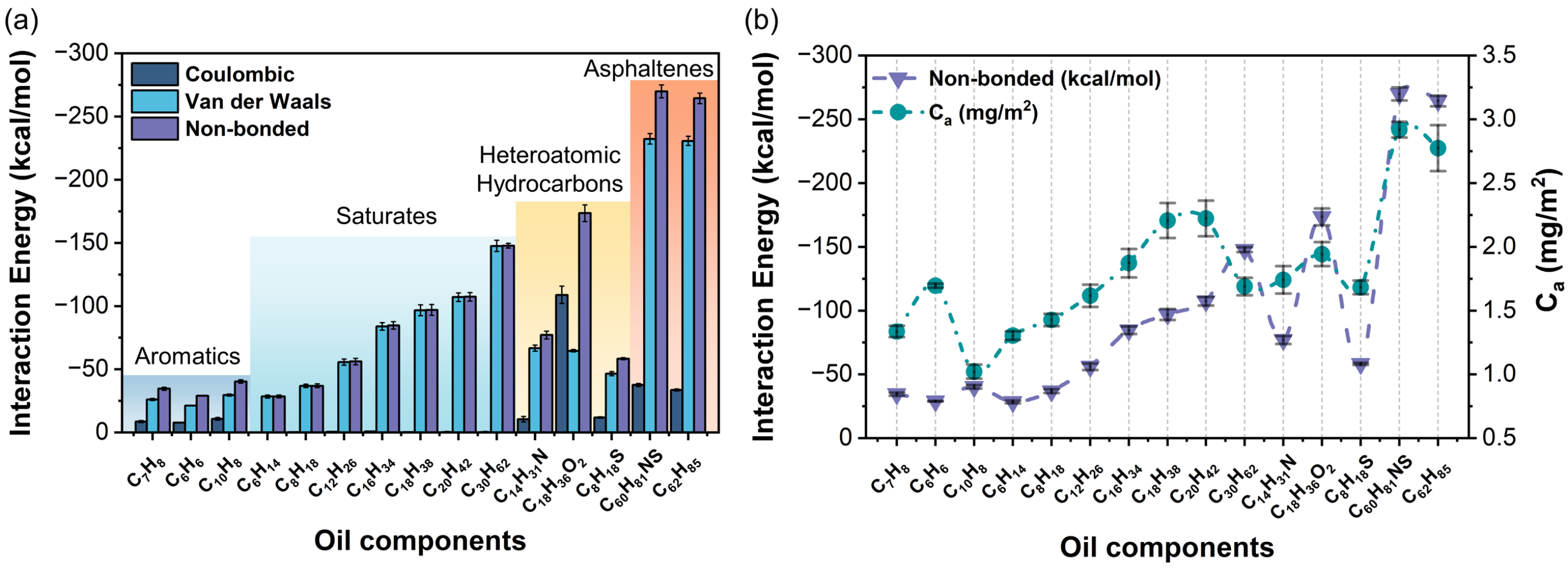

***Fig.7.*** *Interaction energy between different isolated monomer hydrocarbons and kaolinite (a) histograms showing the distribution of interaction energies between different hydrocarbons and clay, where the dark blue area represents aromatic hydrocarbons, the light blue area represents saturated hydrocarbons, the yellow area represents heteroatomic hydrocarbons, and the orange area represents asphaltenes; (b) the distribution of the adsorption amount per unit area and the intermolecular interaction energies of different hydrocarbons on the kaolinite surfaces.*

In this work we assessed the differential interaction energies among three hydrocarbon species, each integrating distinct heteroatoms, with the basal surface of clay. The findings reveal that solely for stearic acid does the interaction with the clay basal surface manifest a predominance of Coulombic forces (as depicted in Fig. 7(a)), with the Coulombic forces contributing 63 % to the aggregate non-bonded interaction energy. The intensity of the interaction energies associated with these three hydrocarbons and the clay basal surface appears to be intricately linked to the characteristics of the polar functional groups they possess. Specifically, the carboxyl moiety of stearic acid is predisposed towards establishing robust electrostatic interactions or hydrogen bonds with the hydroxyl groups on the kaolinite's polar surface. This propensity is also evidenced by the vertical alignment of hydrocarbon molecules on polar surfaces, as illustrated in Fig. 4, which underscores the Coulombic-force-centric nature of the interaction between stearic acid and the clay basal surfaces. On the other hand, dimethyldodecylamine and octanethiol, due to the comparatively lower polarity of their amino and thiol groups, are more likely to interact with kaolinite through van der Waals forces originating from their nonpolar segments. Furthermore, upon conducting a comparative analysis of the adsorption quantities and interaction energies pertaining to these three compounds, it was observed that their relative strength trends remained consistent (Fig. 7(b)), which further confirms the correlation between the interaction mechanisms of heteroatomic hydrocarbon compounds and their adsorption amounts.

Based on Eq. (12), the fluctuation of free energy in the *z*-axis direction for isolated hydrocarbon monomers of different components was calculated, as depicted in Fig. 8. A comparison of the free energy changes of saturated hydrocarbons with different carbon chain lengths shows that the free energy potential wells on both sides of the clay surface correspond to the positions of the adsorption layers in the density curves. This suggests that molecules within the potential well regions attain a lower energy and a more stable state, culminating in the formation of adsorption layers. As the distance from the surface extends, the depth of the potential wells of the adsorption layers gradually decreases, and the energy stability becomes increasingly disrupted. There are certain heights of potential barriers between each potential well, indicating that hydrocarbon molecules must overcome a significant energy barrier to enter or leave the adsorption layer, and the height of this barrier decreases with increasing distance from the surface. For instance, a solitary *n*-eicosane molecule requires overcoming a 0.77 kcal/mol energy barrier to exit the 1$^{st}$ adsorption layer. In contrast, a single alkane molecule must cross a 0.27 kcal/mol barrier from the bulk phase to access the 1$^{st}$ adsorption layer. Given that these barrier values significantly exceed thermal energy, once a stable adsorption layer is established on the kaolinite surface, the molecules are confined within these barriers, making their escape challenging.

The free energy curves illustrate that with longer carbon chains, the potential wells deepen, and the energy barrier to cross the adsorption layers rises, elucidating the inherent mechanism behind the higher adsorption of longer-chain saturated hydrocarbons. A comparison of free energy variations across kaolinite's basal surfaces reveals that the depth and height of the free energy potential wells of hydrocarbon molecules on the SI surface exceed those on the HY surface (for *n*-eicosane, the 1$^{st}$ potential well depth is 0.33 vs 0.27 kcal/mol), further suggesting that saturated hydrocarbon molecules are more stably adsorbed near the SI surface, resulting in larger adsorption amounts. The inset in Fig. 8(a) demonstrates that *n*-eicosane molecules align parallel to the surface and adhere to the surface in three adsorption layers on the SI surface, with a strong energy barrier impeding their interlayer movement. Fig. 8(a) dark blue dashed line denotes the free energy change curve for squalane, indicating that its potential well depth corresponding to the adsorption layer is lower than that of *n*-eicosane, especially on the SI surface (0.18 vs 0.33 kcal/mol). This indicates that branching in the molecular configuration has a significant impact on adsorption, also validations our previous observations.

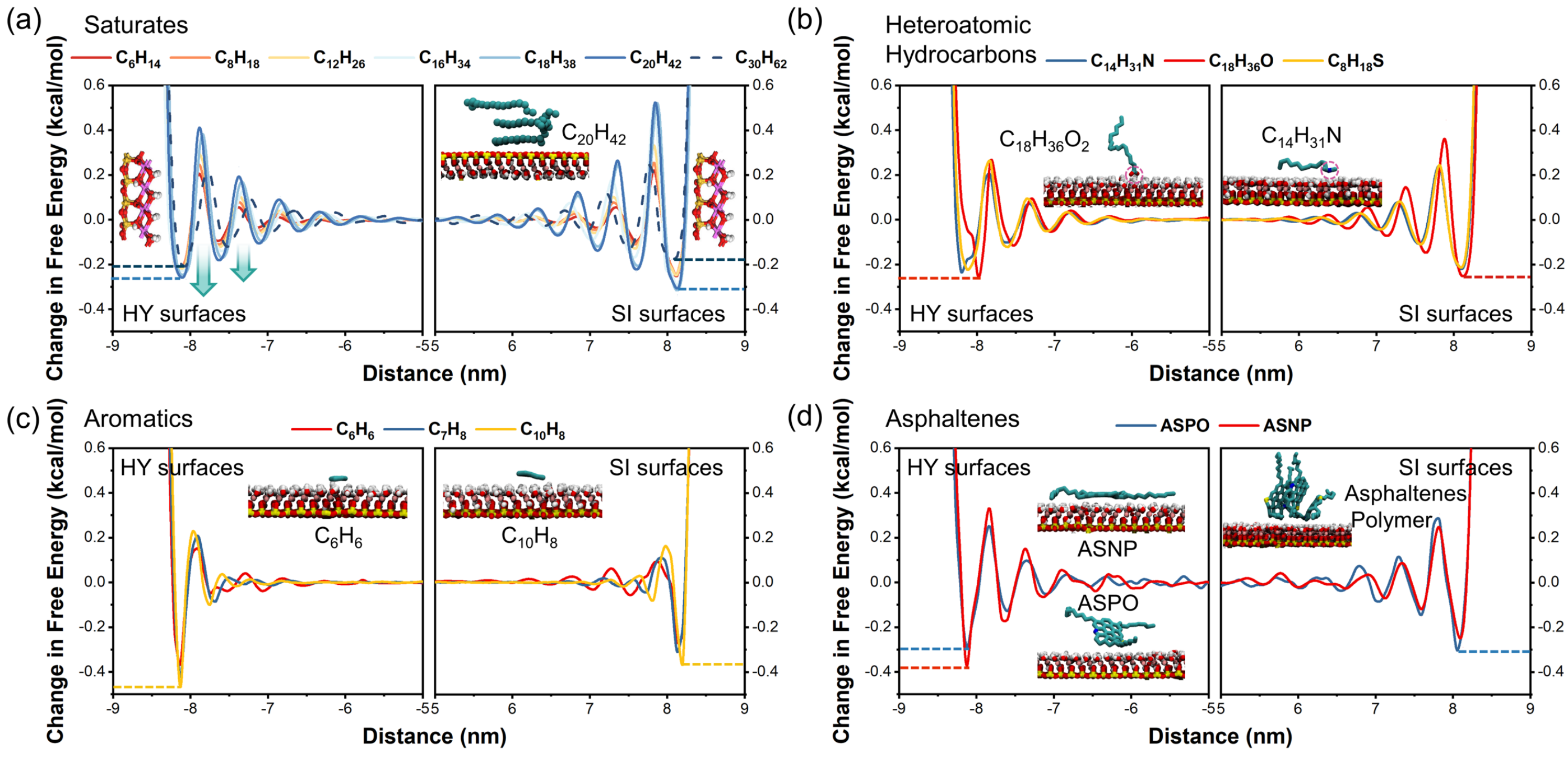

***Fig.8.*** *The distribution of free energy changes with respect to the bulk fluid of isolated monomers hydrocarbons of various components along the z-axis on both surfaces of kaolinite (a) saturated hydrocarbons with different carbon chain lengths, where squalane is represented by a dark blue dashed line, (b) heteroatomic hydrocarbons on both surfaces of kaolinite, with the inset showing the adsorption characteristics of stearic acid and n,n-dimethyldodecylamine on the HY surface, (c) aromatic hydrocarbon molecules, with the inset showing the configuration distribution of benzene and naphthalene on the HY surface; (d) two types of asphaltene molecules, with the inset showing the adsorption characteristics of the two asphaltenes on the HY surface and the polymeric formed by asphaltenes. The inset shows the schematic molecular configuration of both basal surfaces of kaolinite and the adsorpion alignment of individual molecules.*

Fig. 8(b) shows the changes in the free energy for three heteroatomic compounds, elucidating the main differences in their free energy distributions on the polar HY basal surface of kaolinite. Owing to its strong polarity, stearic acid establishes hydrogen bonds with the clay basal surface via its oxygen-rich functional group head, while positioning its non-polar tail perpendicularly to the basal surface (see left inset in Fig. 8(b)). As a result, the free energy curve of stearic acid exhibits a bimodal pattern. The first low well (–0.08 kcal/mol) represents the energy coordination to the HY surface by the carboxylic group, followed by a second deeper well (–0.26 kcal/mol) representing the interaction between the non-polar tail and the surface. This pattern implies a close correlation between the molecule's adsorption configuration and its free energy distribution. The potential wells for nitrogen-containing compounds also show a bimodal shape, yet the nitrogen-containing head functional group of *n,n*-dimethyldodecylamine is less polar than stearic acid, aligns its molecules parallel to the basal surface (as seen in the right inset of Fig. 8(b)). Therefore, its first well is lower, reflecting the hydrogen bond between the nitrogen-containing head and the surface, followed by a higher well denoting the parallel adsorbed non-polar tail. For octanethiol, due to its minimal polarity, its free energy distribution curve is similar to that of saturated hydrocarbons. On kaolinite's non-polar SI basal surface, the heteroatomic compounds are influenced solely by van der Waals interactions and are arranged parallel to the surface, with their free energy change curves consistent with those of saturated hydrocarbons.

Fig. 8(c) shows the free energy change curves for three aromatic hydrocarbons, revealing that the depth of the 1$^{st}$ adsorption layer's potential well for naphthalene surpasses those of benzene and toluene. As depicted in the right inset of Fig. 8(c), naphthalene predominantly adsorbs parallel to the clay basal surface. Moreover, an examination of the potential well depths across both basal surfaces indicates that the instantaneous dipole moments produced by the π-electron clouds of aromatic hydrocarbons facilitate their formation of stable adsorption layers on kaolinite's HY basal surface (for naphthalene, the depth of the 1$^{st}$ adsorption layer's potential well is −0.47 vs −0.35 kcal/mol), thereby necessitating a higher energy barrier to cross these layers. Although the first adsorption layer's potential well depth for benzene is not as deep as those for naphthalene and toluene, its free energy curve shows multilayer adsorption characteristics, which primarily contribute to its greater adsorption amount compared to naphthalene and toluene.

Fig. 8(d) presents the free energy change curves for two types of asphaltene molecules. This indicates that ASNP asphaltene molecules, characterized by their non-polar nature, predominantly adhere parallel to the basal surface (illustrated in the top left inset), resulting in a deeper free energy potential well compared to ASPO asphaltenes that contain heteroatoms (−0.37 vs −0.29 kcal/mol). ASPO asphaltene molecules, containing both nitrogen and sulphur heteroatoms, are inclined to adopt an angled adsorption orientation, with the heteroatoms making contact with the basal surface (depicted in the bottom left inset). This orientation hinders the molecules from attaining a lower energy, more stable state, thereby impacting their overall adsorption capacity. Additionally, owing to the π-π stacking interaction amongst asphaltene molecules, they tend to form dimers or polymers (shown in the right inset), leading to variations in the bulk fluid phase's free energy around zero. On kaolinite's non-polar SI basal surface, van der Waals interactions facilitate a deeper potential well in a stable state for high molecular weight ASPO asphaltenes.

***Table 2***

*Adsorption and interaction energy parameters of isolated monomer hydrocarbons on the kaolinite basal surfaces.*

| Hydrocarbons | $H$ (nm) | $C_a$ (mg/m$^2$) | $\rho_a$ (kg/m$^3$) | $\rho_f$ (kg/m$^3$) | $\rho_{bulk}$ (kg/m$^3$) | $D$ ($10^{-9}$ m²/s) | $E_{Coul}$ (kcal/mol) | $E_{VDW}$ (kcal/mol) | $E_{non\text{-}bonded}$ (kcal/mol) | $E_{Coul}$ : $E_{VDW}$ ratio (% : %) |
|---|---|---|---|---|---|---|---|---|---|---|
| $C_7H_8$ | 2.96 | 1.34 (±0.05) | 902 | 864 | 867 | 1.136 (±0.12) | −8.60 (±0.82) | −25.99 (±0.81) | −34.58 (±1.22) | 25 : 75 |
| $C_6H_6$ | 3.88 | 1.70 (±0.02) | 886 | 867 | 874 | 1.332 (±0.15) | −7.72 (±0.20) | −21.34 (±0.25) | −29.07 (±0.36) | 27 : 73 |
| $C_{10}H_8$ | 1.82 | 1.02 (±0.06) | 1121 | 973 | 963 | 0.493 (±0.01) | −10.75 (±1.01) | −29.54 (±0.85) | −40.30 (±1.38) | 27 : 73 |
| $C_6H_{14}$ | 3.75 | 1.31 (±0.03) | 696 | 641 | 659 | 3.089 (±0.39) | −0.13 (±0.19) | −28.37 (±1.17) | −28.23 (±1.17) | <1 : 99 |
| $C_8H_{18}$ | 3.75 | 1.43 (±0.05) | 762 | 691 | 703 | 1.583 (±0.19) | −0.17 (±0.29) | −36.78 (±1.39) | −37.91 (±1.45) | <1 : 99 |
| $C_{12}H_{26}$ | 3.74 | 1.62 (±0.09) | 866 | 728 | 748 | 0.626 (±0.03) | −0.55 (±0.36) | −55.59 (±2.46) | −56.14 (±2.51) | <1 : 99 |
| $C_{16}H_{34}$ | 3.98 | 1.87 (±0.11) | 941 | 769 | 773 | 0.244 (±0.00) | −0.77 (±0.40) | −83.86 (±2.85) | −83.09 (±2.87) | <1 : 99 |
| $C_{18}H_{38}$ | 4.60 | 2.21 (±0.14) | 960 | 784 | 782 | 0.159 (±0.02) | −0.21 (±0.40) | −96.73 (±4.28) | −96.52 (±4.31) | <1 : 99 |
| $C_{20}H_{42}$ | 4.62 | 2.22 (±0.14) | 963 | 794 | 789 | 0.129 (±0.02) | −0.23 (±0.45) | −107.18 (±3.43) | −106.95 (±3.42) | <1 : 99 |
| $C_{30}H_{62}$ | 4.03 | 1.69 (±0.07) | 838 | 801 | 809 | 0.008 (±0.00) | −0.19 (±0.56) | −147.72 (±4.50) | −147.91 (±1.87) | <1 : 99 |
| $C_{14}H_{31}N$ | 3.72 | 1.74 (±0.11) | 936 | 781 | 787 | 0.292 (±0.01) | −10.44 (±2.10) | −66.62 (±2.41) | −77.07 (±3.05) | 14: 86 |
| $C_{18}H_{36}O_2$ | 4.42 | 1.94 (±0.09) | 980 | 932 | 941 | 0.017 (±0.00) | −108.94 (±6.82) | −64.60 (±0.98) | −173.54 (±6.51) | 63 : 37 |
| $C_8H_{18}S$ | 3.74 | 1.68 (±0.05) | 929 | 837 | 843 | 0.761 (±0.05) | −11.83 (±0.46) | −46.41 (±1.63) | −58.24 (±0.78) | 20 : 80 |
| $C_{60}H_{81}NS$ | 3.70 | 2.92 (±0.06) | 1579 | 920 | - | 0.001 (±0.00) | −37.52 (±1.25) | −232.34 (±4.21) | −269.86 (±5.18) | 14 : 86 |
| $C_{62}H_{85}$ | 3.86 | 2.77 (±0.18) | 1438 | 932 | - | 0.001 (±0.00) | −33.67 (±0.80) | −230.72 (±3.66) | −264.39 (±4.18) | 13 : 87 |

*Where H is the total adsorption thickness, nm; $C_a$ is the amount of adsorption per unit area, mg/m$^2$; $\rho_a$ is the average density of the adsorbed phase, in kg/m$^3$; $\rho_f$ is the average density of the free phase, in kg/m$^3$; $\rho_{bulk}$ is the experimentally measured densities, in kg/m$^3$; D is self-diffusion coefficient, m²/s; $E_{coul}$ and $E_{VDW}$ represent the long-range intermolecular Coulomb and van der Waals interaction terms, respectively. Values in brackets give the standard error as obtained by evaluating outcomes of parallel simulations for different hydrocarbon components.*

### 3.2 The temperature effects on the adsorption features in pure hydrocarbon systems

#### 3.2.1 Adsorption and molecular alignment distributions

In this part of the work, we investigate the changes in adsorption characteristics of *n*-dodecane on the basal surfaces of kaolinite with the change of temperature. Fig. 9(a) and (b) illustrate the hydrocarbon mass density on kaolinite's basal surfaces at a constant pressure (0.1 MPa) across a temperature from 298 K to 453 K. These figures also show the molecular configuration traits at the lowest and highest temperatures. Furthermore, we calculated the bidimensional density distribution probabilities of dodecane at varying temperatures (Fig. 9(c)). With an increase of temperature above ambient conditions (298 K), the mass density fluctuation amplitude of dodecane gradually decreases, the number of adsorption layers lowers, and the peak of adsorption density steadily diminishes until the stabilisation above 373 K (Fig. 9 (a)). Fig. 9 (b) shows that, within the temperature bracket from 373 K to 453 K, the amount of dodecane adsorption layers remains constant, and the density peak value exhibits relative stability. At an environmental temperature of 363 K, the adsorptive characteristics and fluidic behaviour of hydrocarbons on the basal surface of kaolinite remain constant.

From the density distribution and molecular configuration of dodecane at 298 K in Fig. 9 (a), it is evident that lower temperatures facilitate the formation of a wax-like structure in hydrocarbons. The straight-chain saturated hydrocarbon molecules are neatly aligned and form a crystal-like structure along the parallel direction of the kaolinite surface. Notably, this temperature is higher than the melting point of dodecane (i.e., 263 K), which is corroborated by the presence of a liquid-state fluid in the center of the system. This wax-like structure is governed by the intermolecular forces among alkane molecules. Two types of groups comprise alkanes: the terminal $CH_3$ and the primary body $CH_2$. In its fully extended state, the sp3-hybridisation of carbon atoms forms a zigzag configuration. The aggregate intermolecular forces amid $CH_2$ groups surpass those involving the terminal $CH_3$, thus alkanes tend to form orderly arrayed wax-like structures at reduced temperatures and elevated pressures [101]. Under such conditions, hydrocarbons in clay pores tend to develop wax-like formations, impeding flowability, and obstructing pore spaces. With increasing temperature, these crystal-like structures swiftly diminish and the molecular kinetic energy escalates with an increase in temperature. Eventually, reaching a critical threshold, the behaviour of organic hydrocarbon fluids stabilises, persistently exhibiting a coexistence pattern of adsorbed and free states.

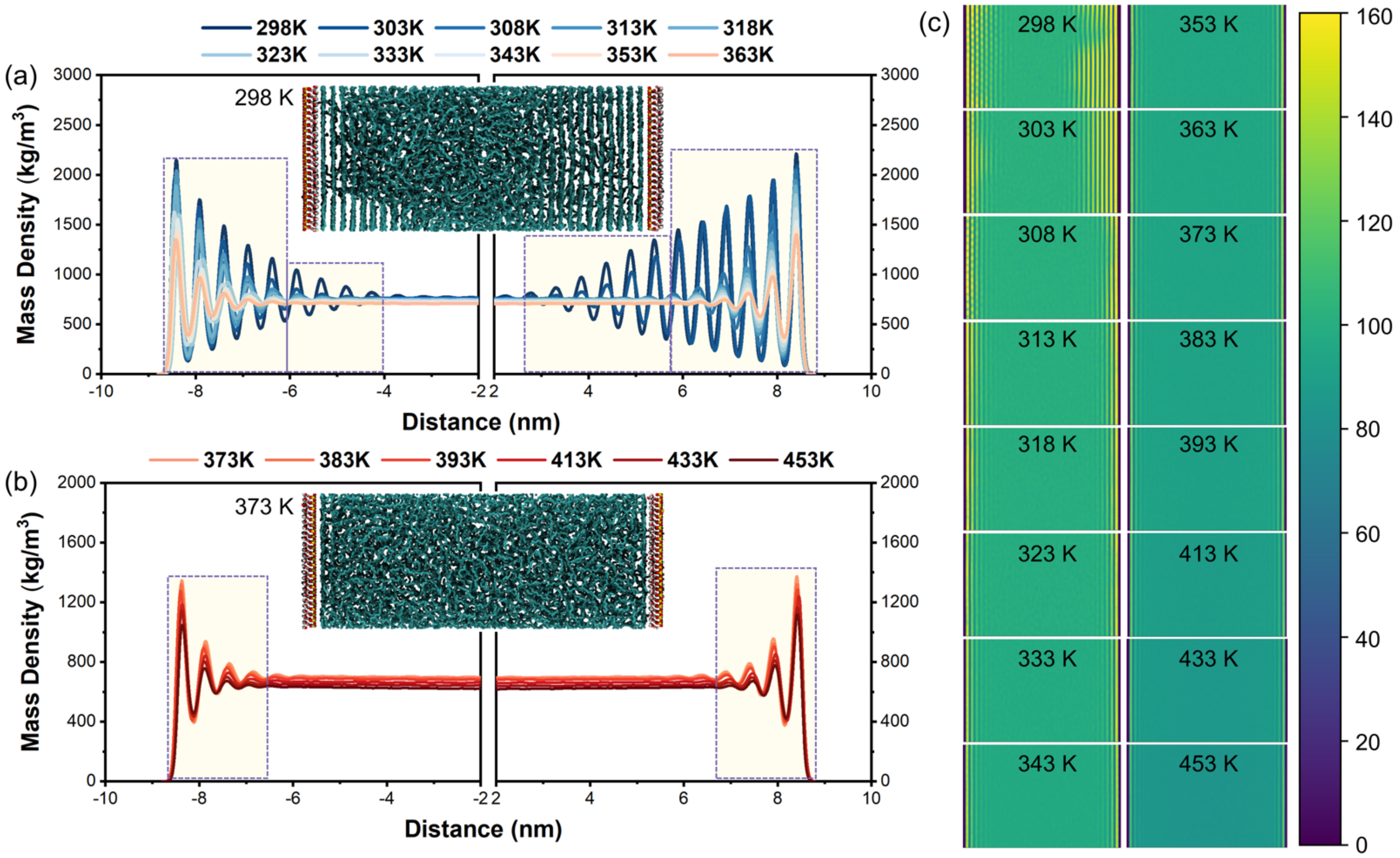

***Fig.9.*** *Distribution of density of n-dodecane on the basal surfaces of kaolinite at varying temperatures at a constant pressure (0.1 MPa) (a) density variation of n-dodecane within the temperature range of 298 K to 363 K, where the inset shows the wax-formation phenomenon of alkanes at 298 K. The yellow area represents the region where molecules are adsorbed; (b) density variation of n-dodecane within the temperature range of 373 K to 453 K, with the inset showing the molecular configurations of alkanes in the kaolinite pore at 373 K; (c) 2D density distribution of n-dodecane in xy-plane along the z-direction at different temperatures.*

### 3.2.2 Changes in fluid dynamics and interaction energies with temperature

At various temperatures, the adsorption of *n*-dodecane per unit area of kaolinite, the diffusion coefficient, and the interaction energy between hydrocarbons and clay were calculated to elucidate the critical influence of temperature on the fluid behaviour of hydrocarbons within kaolinite's pores. The fluctuation in dodecane adsorption on the basal surface of kaolinite with increasing temperature is depicted in Fig. 10(a). The data indicate a significant decline in adsorbed species as the temperature rises. At 298 K, attributed to the wax-formation phenomenon induced by pressure, the adsorption amount of dodecane reaches 3.87 mg/m$^2$. As the temperature increases to 323 K, the adsorption decreases markedly (i.e., 1.64 mg/m$^2$). At 363 K and above, adsorption behaviour no longer changes, maintaining an average of 1.25 mg/m$^2$. Fig. 10(b) shows the alteration in the mean value of the monolayer adsorption thickness of hydrocarbons with increasing temperature, where the adsorption thickness exhibits a linear upward trend with elevation of the temperature. This trend suggests that the rise in temperature enhances the kinetic energy of hydrocarbon molecules. Consequently, due to their interaction with the clay surface, the molecules become adsorbed and confined within a specific area. Once a certain temperature threshold is surpassed, the adsorption ceases to fluctuate, signifying that at this juncture, the dodecane molecules have achieved a dynamic equilibrium between adsorption and desorption.

Fig. 10(c) illustrates the changes in the molecular diffusion coefficient with temperature increase. The increase in the self-diffusion coefficient of molecules is directly proportional to the rise in molecular kinetic energy with temperature. At lower temperatures, the mobility of hydrocarbons within clay pores is markedly low (approximately $0.21\times10^{-9}$ $m^2/s$) due to the wax-formation phenomenon at the interface. On the contrary, as the temperature increases to 453 K, the diffusion coefficient of dodecane increases substantially to $3.69\times10^{-9}$ $m^2/s$. This trend implies that a rise in temperature significantly boosts the mobility of hydrocarbons, and the waxing phenomenon rapidly diminishes with temperature increase. The scope of this study encompasses temperatures under geological conditions, suggesting that temperature evolution in geological environments profoundly impacts the adsorptive behaviour of hydrocarbons. Nevertheless, once the temperature attains a certain threshold, the adsorptive capacity of the clay basal surface for hydrocarbons becomes lower thermal sensitivity. This work utilised *n*-dodecane, identifying a temperature threshold of 363 K. However, it is noteworthy that this threshold may vary depending on factors such as the properties of the basal surface and hydrocarbon components.

Moreover, the interaction energy between the kaolinite basal surface and dodecane was computed to elucidate the micro-mechanisms influenced by temperature. Fig. 10(c) depicts the evolution of interaction energy as temperature surges, revealing a 40 % decrease in interaction energy when the temperature increases from 298 K to 453 K. This observation indicates that, as the temperature increases, the adsorption affinity of dodecane molecules for the basal surface of kaolinite steadily diminishes. The increased molecular motion at elevated temperatures results in increased collisions and thermal vibrations among molecules, curtailing their residency on the clay basal surface. Concurrently, the temperature rise also activates the adsorption and desorption processes of dodecane molecules, expediting the attainment of dynamic equilibrium and thereby reducing the number of molecules adsorbed on the kaolinite surface.

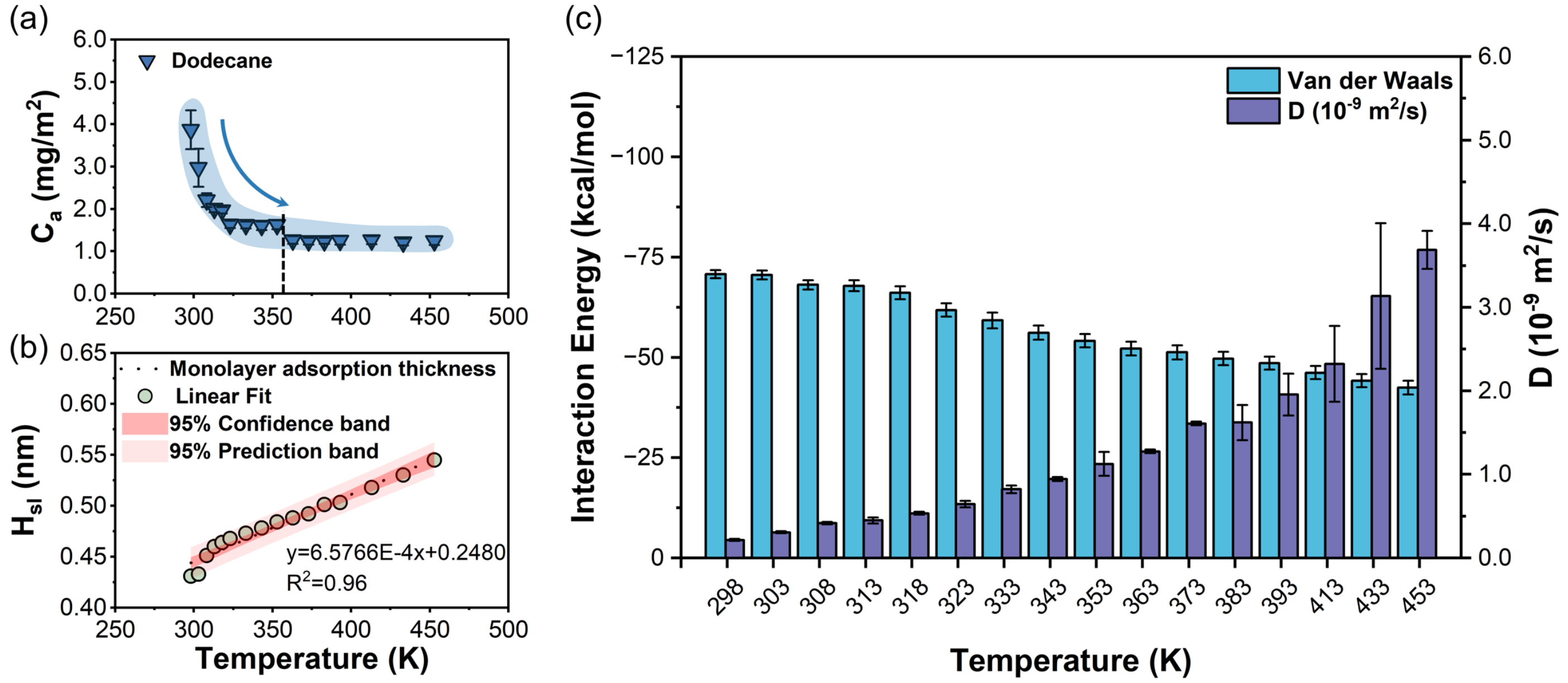

***Fig.10.*** *Adsorption, diffusion characteristics and interaction energy evolution of dodecane on the kaolinite basal surfaces across temperature range. (a) Variation of the adsorbed amount of n-dodecane per unit area with temperature; (b) change in the average adsorption thickness of a single layer of alkanes with temperature; (c) histograms showing the variation of interaction energy between n-dodecane and kaolinite and the self-diffusion coefficient of alkanes with temperature.*

### 3.3 The pressure effects of the adsorption features of pure hydrocarbons

In addition to the effects of temperature on the adsorption of the alkanes on kaolinite basal surfaces, we also evaluate the effect of pressure. Fig. 11(a) and (b) exhibit the mass density distribution of dodecane on both kaolinite surfaces as the pressure increases from 0.5 MPa to 60 MPa in 13 intervals at a constant temperature (323 K), along with the visualisation of the molecular configuration of dodecane at both 0.5 MPa (Fig 11(a)) and 60 MPa (Fig 11(b)). Fig. 11(a) demonstrates that the count of adsorption layers and the peak values of adsorption density remain relatively constant within the pressure spectrum of 0.5 MPa to 40 MPa. This suggests that under pressures below 40 MPa, the effect of pressure on the fluid behaviour of hydrocarbons within clay is negligible. Nevertheless, with a further increment in pressure, alterations occur in the density distribution of hydrocarbon molecules, particularly on the polar hydroxylated surface of kaolinite. As depicted in Fig. 11(b), upon reaching 45 MPa, the peak density of dodecane on the hydroxylated surface intensifies, and the number of adsorption layers increases, while the density curve on the non-polar siloxane basal surface exhibits minimal change. The molecular configuration of dodecane at 60 MPa reveals that on the hydroxylated surface, molecules align in more parallel layers to the surface compared to atmospheric pressure conditions.

The bidimensional density distribution of dodecane, illustrated in Fig. 11(c), more distinctly portrays the adsorption layers formed under elevated pressure, indicating that an increase in pressure fosters a more organised arrangement of hydrocarbons within the system and leads to the development of additional adsorption layers due to spatial constraints.

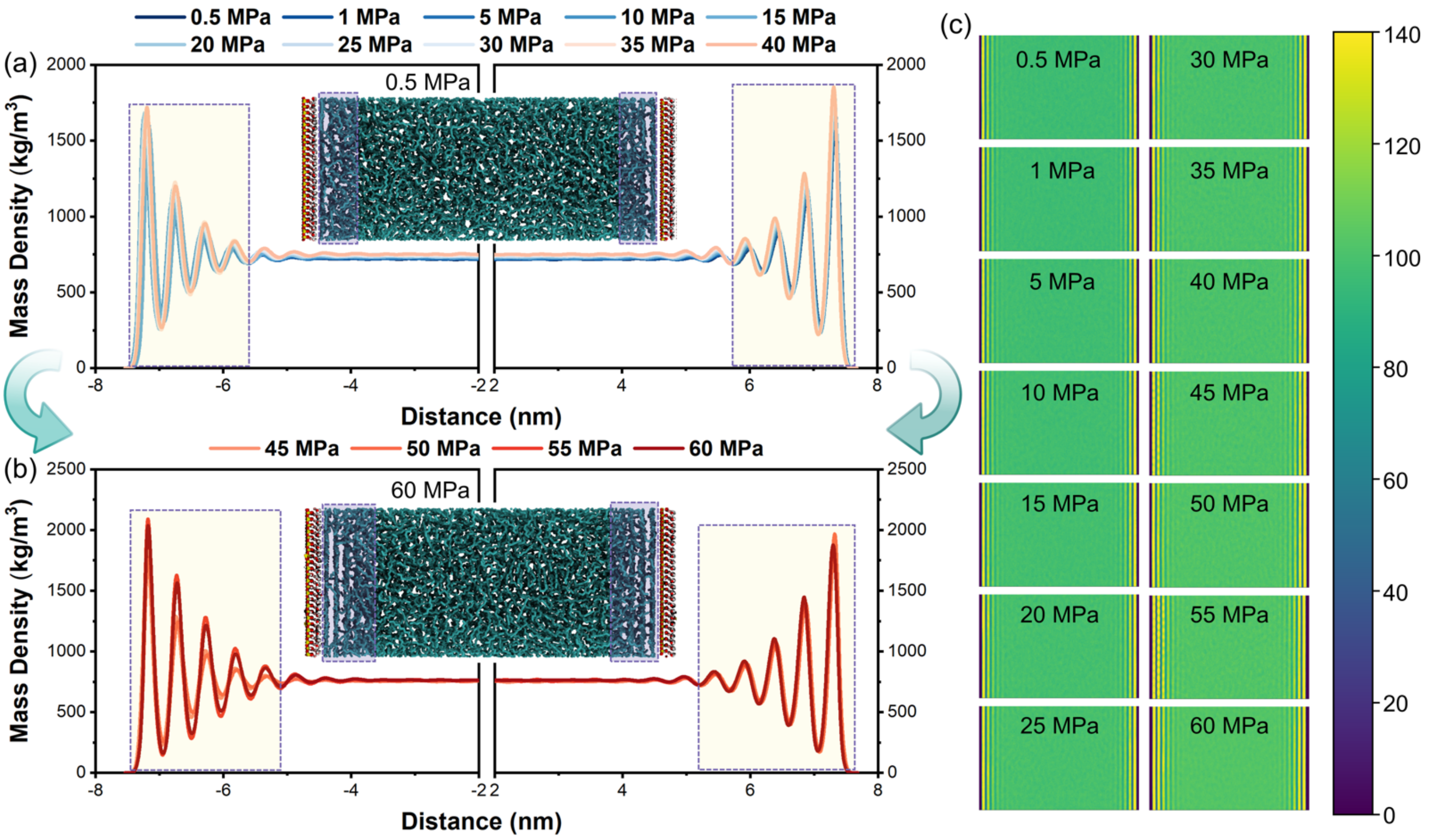

***Fig.11.*** *Distribution of density and molecular configuration of dodecane on the basal surfaces of kaolinite across pressure range at a constant temperature (323 K) (a) Density variation of n-dodecane within the pressure range of 0.5 MPa to 40 MPa, where the inset shows the molecular configuration of alkanes at low pressure of 0.5 MPa. The yellow area represents the region where molecules are adsorbed, and the purple area represents the corresponding region in the molecular configuration where molecules are adsorbed; (b) density variation of n-dodecane within the pressure range of 45 MPa to 60 MPa, with the inset showing the molecular configuration of alkanes at 60 MPa; (c) 2D density distribution of n-dodecane in xy-plane along the z-direction under different pressures.*

The objective of the statistical analysis of molecular orientation is to clarify how pressure influences the interaction between molecules of *n*-dodecane and the basal surfaces of kaolinite. Fig. 12(a) and (b) illustrate the distribution disparities in the elevation and azimuth angles of dodecane molecules under pressures of 0.5 MPa and 60 MPa. Observations indicate that within the adsorption layers, dodecane molecules predominantly align parallel to the surface, chiefly orienting along the hexagonal symmetry axis of the siloxane surface. Fig. 12(a) shows that on both sides of the kaolinite surfaces, dodecane molecules constitute four adsorption layers, albeit with considerable dispersion at azimuth angles. Conversely, at a pressure of 60 MPa, the number of dodecane adsorption layers escalates to five, exhibiting enhanced uniformity in azimuth angles (Fig. 12(b)). This underscores that a higher pressure induces a more structured arrangement of hydrocarbon molecules in the adsorption layers, culminating in additional adsorption layers.

Fig. 12(c) presents the diffusion coefficient of dodecane and its interaction energy with the kaolinite basal surface as pressure varies. The findings reveal that as pressure increases, the self-diffusion coefficient of dodecane molecules decreases while their energy of interaction with clay augments. This suggests that an increase in pressure inhibits the mobility of hydrocarbon molecules within confined pore volumes and enhances their affinity for the surface. From a thermodynamic perspective, the pressure dependence of hydrocarbons' adsorption correlates with the volumetric change of the system. It was discerned that within an equilibrated system, individual alkane molecules in the bulk fluid occupy more volume than alkanes methodically arrayed in the adsorption layers. Therefore, as pressure increases, the space accessible for molecules decreases, resulting in a diminished capacity for self-diffusion. This, in turn, causes an increase in both the number of adsorption layers and the adsorption amount. Fig. 12(d) shows a graph of the adsorption amount as a function of pressure, indicating that a minor increase in initial pressure does not substantially affect the adsorption amount of hydrocarbons. However, beyond a certain pressure threshold (45 MPa), the amount of adsorption significantly escalates, closely associated with the formation of additional adsorption layers.

In conclusion, this research substantiates the hypothesis that in subterranean high-pressure environments or under ultra-high-pressure conditions, the influence of pressure on adsorption at the solid-liquid interface is considerable and cannot be overlooked. This investigation enriches the understanding of the adsorption behaviour of organic hydrocarbon fluids in deep high-pressure strata and facilitates the evaluation of interaction characteristics between organic hydrocarbon fluids and clay basal surfaces under extreme conditions. At a specific pressure threshold, the amount of hydrocarbon adsorption on the basal clay surfaces may experience a marked change. This occurrence, coupled with pore blockage due to the diminished mobility of hydrocarbon molecules, can adversely impact the recovery efficiency of crude oil.

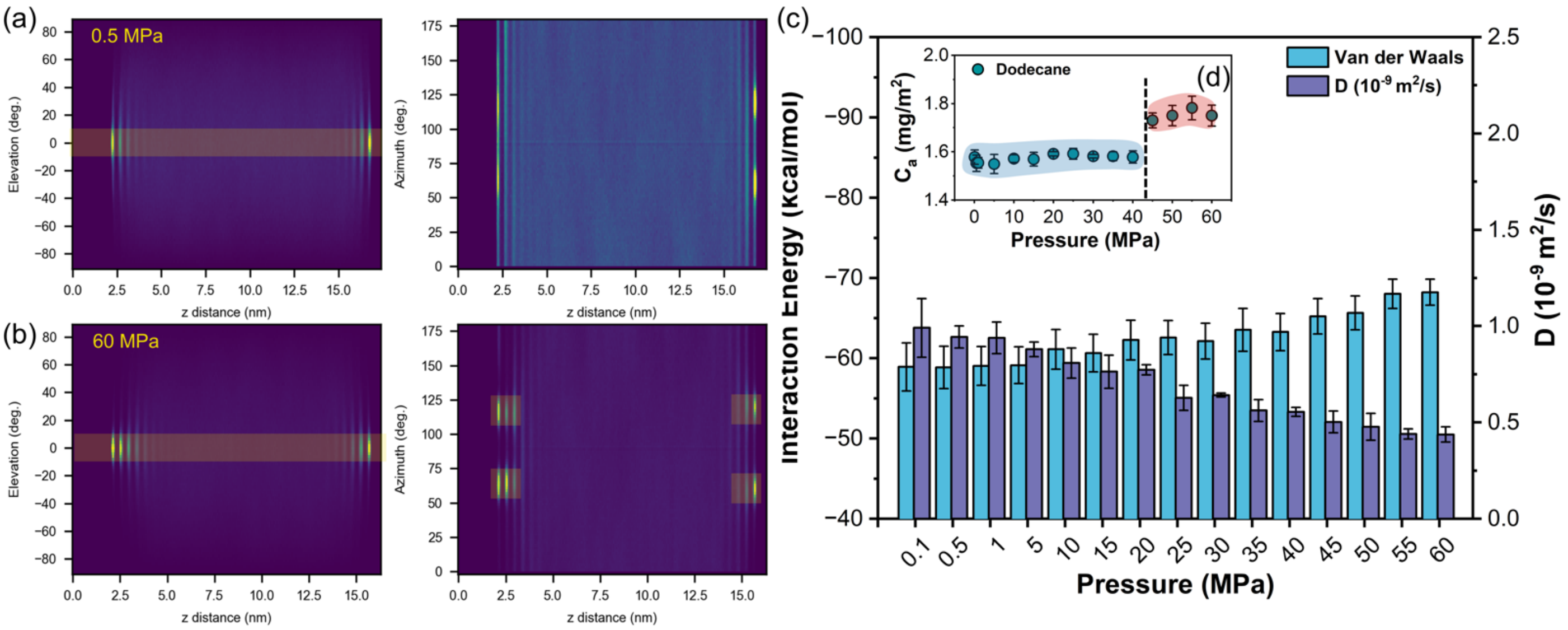

***Fig.12.*** *Molecular orientations, adsorption, diffusion characteristics and interaction energy evolution of dodecane on the kaolinite basal surfaces at different pressure (a) orientation characteristics of n-dodecane on both surfaces of kaolinite at 0.5 MPa; (b) orientation characteristics of alkanes at high pressure of 60 MPa; (c) histograms showing the variation of interaction energy between n-dodecane and clay as well as the self-diffusion coefficient of alkane molecules with pressure, where the inset illustrates the evolution of the adsorption amount of alkanes per unit area with increasing pressure.*

### 3.4 Adsorption behaviour of mixed-component hydrocarbons on kaolinite surfaces

In this section, we discuss the adsorption behaviour of multi-component oil mixture on kaolinite surfaces. After 400 ns of NPT ensemble simulation and 200 ns of periodic temperature annealing simulation, the equilibrium molecular configuration and density distribution of components in multicomponent hydrocarbon systems were analysed. This analysis aimed to elucidate the interactions and competitive adsorption effects among various hydrocarbons under geological conditions. The molecular configuration features at equilibrium are illustrated in Fig. 13(a). The left side of the kaolinite illustrates its polar HY surface, while the right side represents the non-polar SI surface. The molecular configuration analysis highlights marked adsorption disparities for hydrocarbon components on the kaolinite basal surface, uncovering significant competitive adsorption behaviour among hydrocarbons. As illustrated in Fig. 13(a), the polar surface of kaolinite is predominantly occupied by heteroatomic hydrocarbons, obscuring the adsorption of other hydrocarbon components on this surface. On the contrary, parallel-arranged adsorption layers of hydrocarbons are observable on the siloxane surface. The primary layer consists of alkanes and asphaltenes, which are orientated in a parallel direction to the surface.

Based on the amount of carbon atoms, the straight-chain saturated hydrocarbons of mixed components are classified as light or heavy saturated hydrocarbons in this study, with n-dodecane serving as the demarcation line. In order to enhance the comparability of competitive adsorption characteristics among various hydrocarbon components, normalization was applied to the number density distributions of hydrocarbons for five components. The monomolecular probability density distributions of five distinct hydrocarbon components on both sides of kaolinite were acquired via this method, as seen in Fig. 13(b). It can be seen that on the HY surface of kaolinite, heteroatomic hydrocarbons are primarily adsorbed (accounting for 13.26 % of the total adsorption on that surface). Their density curves show an irregular bimodal distribution, indicating that the molecules are not adsorbed in a parallel arrangement. As shown in Fig. 13(a), heteroatomic hydrocarbons on the polar surface of kaolinite are mainly adsorbed in a vertical state, a phenomenon attributed to the hydrogen bonding interactions formed between their functional groups (such as carboxyl and amino groups) and the hydroxyl groups on the clay surface, consistent with the simulation results of isolated monomeric hydrocarbons (see Fig. 4). The 2D density distribution in Fig. 13(c) also shows the primary adsorption of heteroatomic hydrocarbons on the HY surface. The analysis of the probability density curves reveals that aromatic hydrocarbons adsorb substantially in the 1$^{st}$ adsorption layer on the polar surface, constituting 8 % adsorption on the surface. This finding is consistent with the outcomes of the simulations conducted on monomeric hydrocarbons. Nevertheless, the HY surface of kaolinite has a tendency to adsorb heteroatomic hydrocarbons that possess polarity due to competing adsorption processes. This strong adsorption effect also impedes the adsorption of other organic hydrocarbon components on that surface. The effect of amphiphilic compounds (which possess surfactant qualities) in facilitating the desorption of additional organic hydrocarbon components from clay surfaces can be elucidated by this occurrence. Meanwhile, evidence suggests that competing adsorption effects considerably impede the adsorption of heavy components, such as asphaltenes, onto polar surfaces in systems containing heteroatom hydrocarbon molecules.

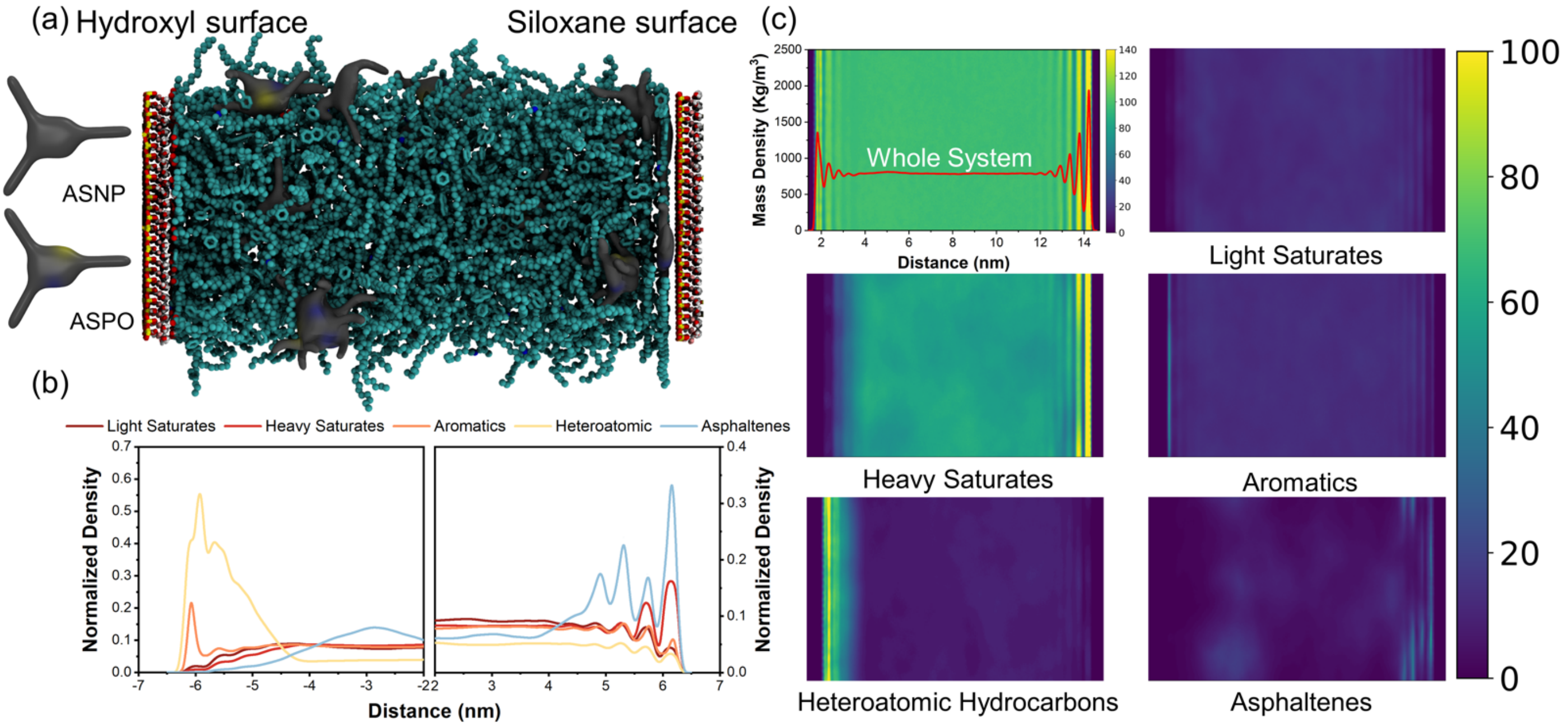

***Fig.13.*** *Molecular configurations and density distribution characteristics of different fractions in multicomponent hydrocarbon mixtures (a) the properties of molecular configuration distribution following equilibrium in annealing simulations, in which hydrogen atoms are removed to clarify the distribution of various component molecules in the system. The isosurface representation (QuickSurf rendering option in the VMD software) is utilized to depict two forms of asphaltenes in grey, highlighting their distribution throughout the mixed system; (b) normalized density distribution of different components on both basal surfaces of kaolinite; (c) 2D density distribution of the system and of each individual hydrocarbon component within the system.*

On the right-hand side of Figure 13(b), the hydrocarbon probability density curve on the SI surface demonstrates distinct stratified adsorption properties. Studies have calculated that the non-polar surface of kaolinite is mainly occupied by heavy saturated hydrocarbons, accounting for 35 % of the total adsorption on the SI surface, with normal alkane molecules forming multilayer adsorption structures parallel to the surface (as shown in Fig. 13(c)). This is consistent with previous findings (see Fig. 4), indicating that long-chain saturated hydrocarbons tend to form regularly arranged adsorption layers parallel to the non-polar surface of kaolinite. The presence of these adsorption layers restricts the capacity of additional components to adsorb onto this surface and reduces the volume of available pore space.

Additional examination demonstrates that asphaltenes, comparable to alkanes, have a tendency to establish adsorption patterns parallel to the surface on the SI surface of kaolinite. This can be attributed to the polycyclic aromatic characteristic of asphaltenes. The mixed components utilized in this study contain asphaltenes in a small amount of mass proportion; hence, their influence is restricted and fails to substantially impede the adsorption of the remaining components. On the contrary, asphaltenes are predominantly adsorbed at the 1st layer of adsorption sites in close proximity to the surface, as indicated by the isolated monomers hydrocarbon probability density distribution. This suggests that an augmentation in asphaltene concentration might potentially impact the adsorption of other hydrocarbons on the SI surface through a competitive adsorption mechanism. Light saturated hydrocarbons with lower carbon numbers mostly exist in a free state in the bulk fluid phase due to competitive adsorption effects (as shown in Fig. 13(c)).

### 3.5 Molecular self-assembly mechanism of mixed hydrocarbons

To illustrate the self-assembly process of organic hydrocarbon molecules in kaolinite pores, images depicting the molecular configuration of different components at different time points within the system were obtained using trajectory files achieved from periodic annealing simulations (as shown in Fig. 14). According to this figure, light saturated hydrocarbons may transiently adsorb onto the clay surface during the equilibration process, but they predominantly exist freely in the bulk fluid phase, largely due to the competitive adsorption of other components. On the other hand, heavy saturated hydrocarbons initially occupy adsorption sites on the non-polar surface and eventually develop multi-layered adsorption structures on this surface, stabilizing by the end of the periodic annealing process (160 ns). Aromatic hydrocarbons quickly adhere to the polar surface at the outset, demonstrating parallel configurations on both clay surfaces at 60 ns and 140 ns. However, as the system nears equilibrium, each side of the clay surface becomes progressively dominated by heavy saturated and heteroatomic hydrocarbons, facilitating the desorption of aromatic hydrocarbon molecules.

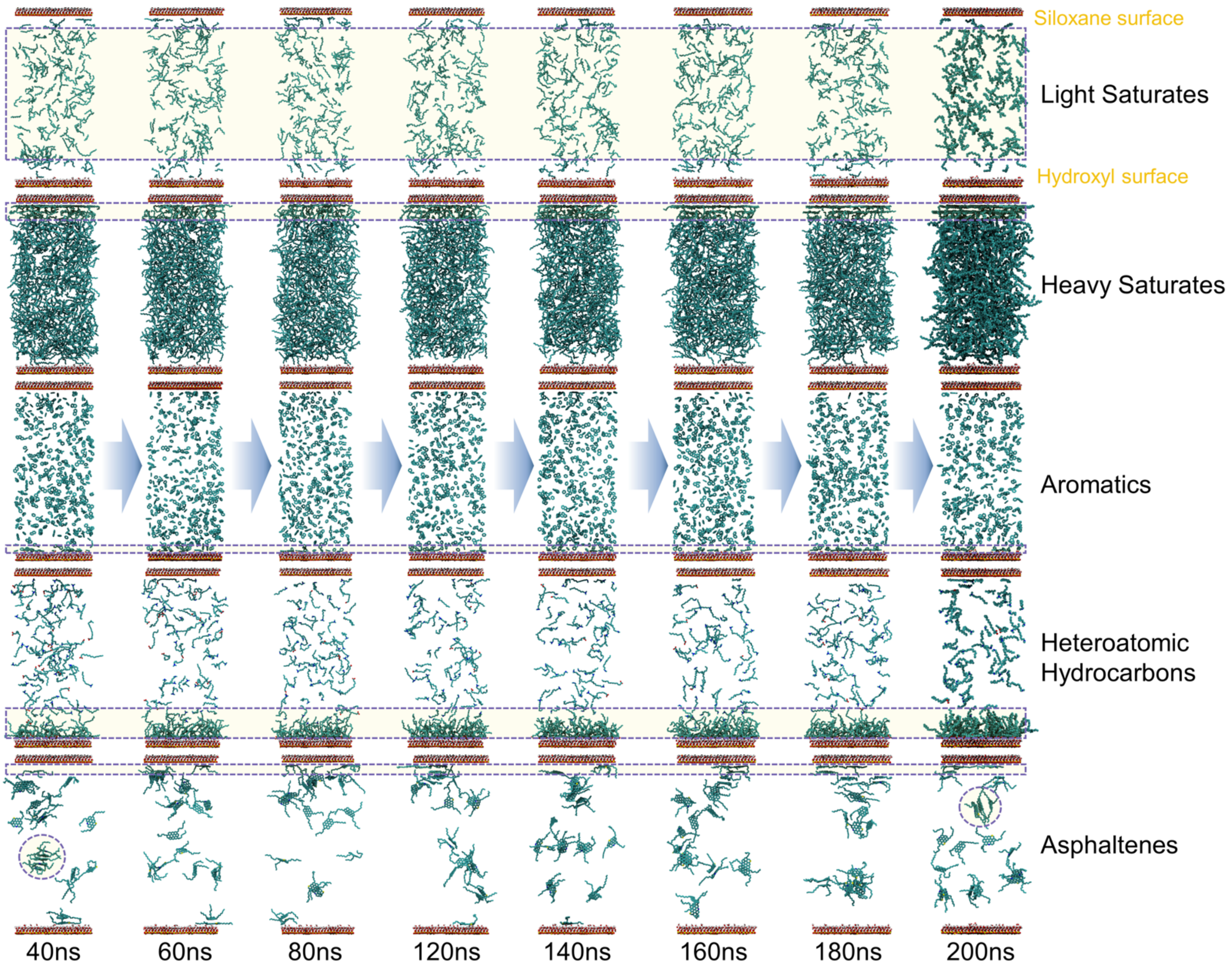

***Fig.14.*** *Dynamical self-assembly process of different component organic hydrocarbon molecules in a mixed system.*

The self-assembly process of heteroatomic hydrocarbons, as depicted in Fig. 14, indicates that these molecules interact primarily vertically with the polar clay surface through their functional groups. Throughout the progression of the simulation towards equilibrium, their adsorption on the surface gradually intensifies, and a minority of these components are observed to have parallel adsorption to the siloxane surface of kaolinite. The self-assembly of asphaltenes presents a more intricate pattern. Initially, many asphaltene molecules, due to π-π interactions between their polycyclic aromatic structures, form asphaltene polymers through π-π stacking, while a few adsorb parallel to the polar surface. As the simulation extends to 120 ns, the self-aggregation of asphaltenes often disrupts their compact adsorption structure, enabling other light hydrocarbon molecules to facilitate their desorption from the clay surface, which is consistent with prior study [102]. When equilibrium is reached, the formation of asphaltene polymers increases. However, their large molecular weight and volume make asphaltenes less prone to adsorption on the clay surface. Only a limited number of asphaltenes that do not form polymers exhibit parallel adsorption to the non-polar surface of kaolinite. This suggests that asphaltenes interact primarily with the clay surface via van der Waals forces, corroborating the simulation findings of isolated monomers.

## 4. Conclusion

Molecular dynamics simulations were used to investigate the adsorption behaviour and fluid dynamics characteristics of an array of hydrocarbons on two distinct basal surfaces of kaolinite clay. The study highlighted the effects of molecular structure, functional groups, and environmental factors (including temperature and pressure) on the adsorption density, molecular configuration, orientation, and mobility of hydrocarbons in nanopore confinement. Additionally, the study clarifies the microscale mechanisms through an examination of intermolecular interaction energies. The findings are summarized as follows:

1) Marked disparities are observed in the fluid behaviours of diverse hydrocarbon components on siloxane (SI) and hydroxyl (HY) basal surfaces of kaolinite. Straight-chain saturated hydrocarbons exhibit a propensity to align parallel and adsorb onto the SI surface of kaolinite, with the amount of adsorption increasing as the carbon chain lengthens. This phenomenon correlated with the amplification in molecular polarizability and the increase of dispersion interaction forces. As the adsorption quantity of saturated hydrocarbons increases, there is a discernible decrease in the self-diffusion coefficient, indicative of diminished mobility. In contrast, aromatic hydrocarbons predominantly adsorb on the polar HY surface, driven by electrostatic interactions. Straight-chain saturated hydrocarbons and benzene demonstrate a tendency to form multilayer adsorptive structures parallel to the kaolinite surface, resulting in heightened adsorption quantities compared to branched molecules, such as squalane and toluene. This highlights the impact of molecular structure and spatial effects on adsorption behaviour.
2) The adsorption capacity of heteroatomic hydrocarbons on kaolinite's basal surfaces manifests a descending sequence of oxygen, nitrogen, and sulphur-containing compounds, consistent with the electronegativity of these elements within their functional groups. Stearic acid, with its amphiphilic nature, displays a dual-orientation molecular alignment on different sides of the clay surface. It predominantly adsorbs in a configuration perpendicular to the polar HY surface, which is related to the formation of hydrogen bonds and electrostatic interactions. The diffusion coefficients of heteroatomic hydrocarbons inversely proportional to their adsorption amount, with octanethiol exhibiting the strongest diffusion capability, while stearic acid remains virtually immobile within kaolinite's pores. The study indicates that the depth of free energy potential wells of hydrocarbons on the kaolinite surface corresponds to the locations of their adsorption layers, indicating that molecules within these layers are in a stable and low-energy state. From an energetic standpoint, the interaction energy between asphaltenes and kaolinite surfaces is the highest, followed by heteroatomic compounds, and then by saturated and aromatic hydrocarbons. Asphaltenes predominantly exhibit multi-layer adsorption characteristics on clay surfaces via van der Waals interactions. Heteroatom-containing asphaltenes show a pronounced affinity for the HY surface, attributable to local electronegativity variations induced by heteroatoms in the aromatic rings.

3) The adsorption properties of hydrocarbons on clay surfaces exhibit notably thermal sensitivity. An increase in temperature precipitate a swift decline in adsorption quantities, correlating with an increase in molecular kinetic energy and a reduction in adsorptive affinity. At lower temperatures (exceeding the crystallization point), dodecane forms a wax-like conformation, but this structure progressively diminishes and stabilizes as temperatures rise. There exists a discernible temperature threshold; surpassing this limit, the fluid behaviour of hydrocarbons within clay pores stabilizes, leading to thermal sensitivity negligible. The fluid dynamics of organic hydrocarbon are mainly unaffected by pressure under a certain threshold. However, when pressure exceeds this limit, the adsorptive properties of the fluids are substantially altered. Increased pressure induces a more structured arrangement of hydrocarbons due to spatial constraints, thereby amplifying their interaction energy with clay. This results in an increase in the number of adsorption layers and adsorption amount, precipitating a substantial reduction in molecular mobility.
4) The simulation results for the system consisting of mixed-components of hydrocarbons, which are established based on their distribution in actual geological environments, reveal significant variations in adsorption among the different surfaces of kaolinite. Influenced by competitive adsorption and interaction with the clay, heteroatomic hydrocarbons have mostly enveloped the polar surface of kaolinite, adopting a perpendicular molecular alignment. Conversely, on the non-polar surface, asphaltenes and heavy saturated hydrocarbons establish multi-layered adsorptive configurations, with molecules oriented parallel to the surface. Notably, asphaltene molecules form polymers through $\pi$-$\pi$ stacking interactions, which destabilizes their tight adsorptive structure and, to a certain degree, facilitates the desorption of hydrocarbon molecules.

These findings contribute to gaining a detailed description of the interactions between hydrocarbons and clay minerals, by analysing the factors that influence the adsorption of hydrocarbons onto the surfaces of clay, contingent upon the characteristics of hydrocarbon components and environmental conditions. Elucidating the fundamental microscopic mechanisms supports the development of more accurate estimations of reservoir fluid distribution characteristics under geological temperature and pressure conditions, as well as in forecasting fluid behaviour. Furthermore, such insights lay the groundwork for revealing the fluid distribution patterns within shale reservoirs and for effectively controlling organic pollutants in the environment.

## Declaration of Competing Interest

The authors declare that they have no known competing financial interests or personal relationships that could have appeared to influence the work reported in this paper.

## Acknowledgements

We would like to express appreciation to the following financial support: National Natural Science Foundation of China (No. 42072160 and No. 42272149), the RIPED (2022-JS-1223), Fundamental Research Funds for the Central Universities (YCX2021002). Rixin Zhao author also acknowledges China Scholarship Council for financial support for a one-year visit to Durham University (No. 202206450073). This work made use of the facilities of the N8 Centre of Excellence in Computationally Intensive Research (N8 CIR) provided and funded by the N8 research partnership and EPSRC (Grant No. EP/T022167/1). The Centre is co-ordinated by the Universities of Durham, Manchester and York.

Supplementary material for

# Revealing crucial effects of reservoir environment and hydrocarbon fractions on fluid behaviour in kaolinite pores

**Rixin Zhao[a, b, c], Haitao Xue[a, b], Shuangfang Lu[d, *], H. Chris Greenwell[c], Valentina Erastova[e, *]**

[a]School of Geosciences, China University of Petroleum (East China), Qingdao 266580, Shandong, China

[b]Key Laboratory of Deep Oil and Gas, China University of Petroleum (East China), Qingdao, 266580, China

[c]Department of Earth Sciences and ⁄⁄Department of Chemistry, Durham University, Durham DH1 3LE, United Kingdom

[d]Sanya Offshore Oil & Gas Research Institute, Northeast Petroleum University, Sanya 572025, China

[e]School of Chemistry, University of Edinburgh, David Brewster Road, Edinburgh, EH9 3FJ, United Kingdom

E-mail addresses: lushuangfang@upc.edu.cn (Shuangfang Lu), valentina.erastova@ed.ac.uk (Valentina Erastova).

**This document contains the following sections:**

Fig. S4: The evolution of system density and the density of each hydrocarbon component over time in simulations under NPT ensemble conditions

Fig. S5: The evolution of system density and the density of each hydrocarbon component over time in periodic annealing simulations

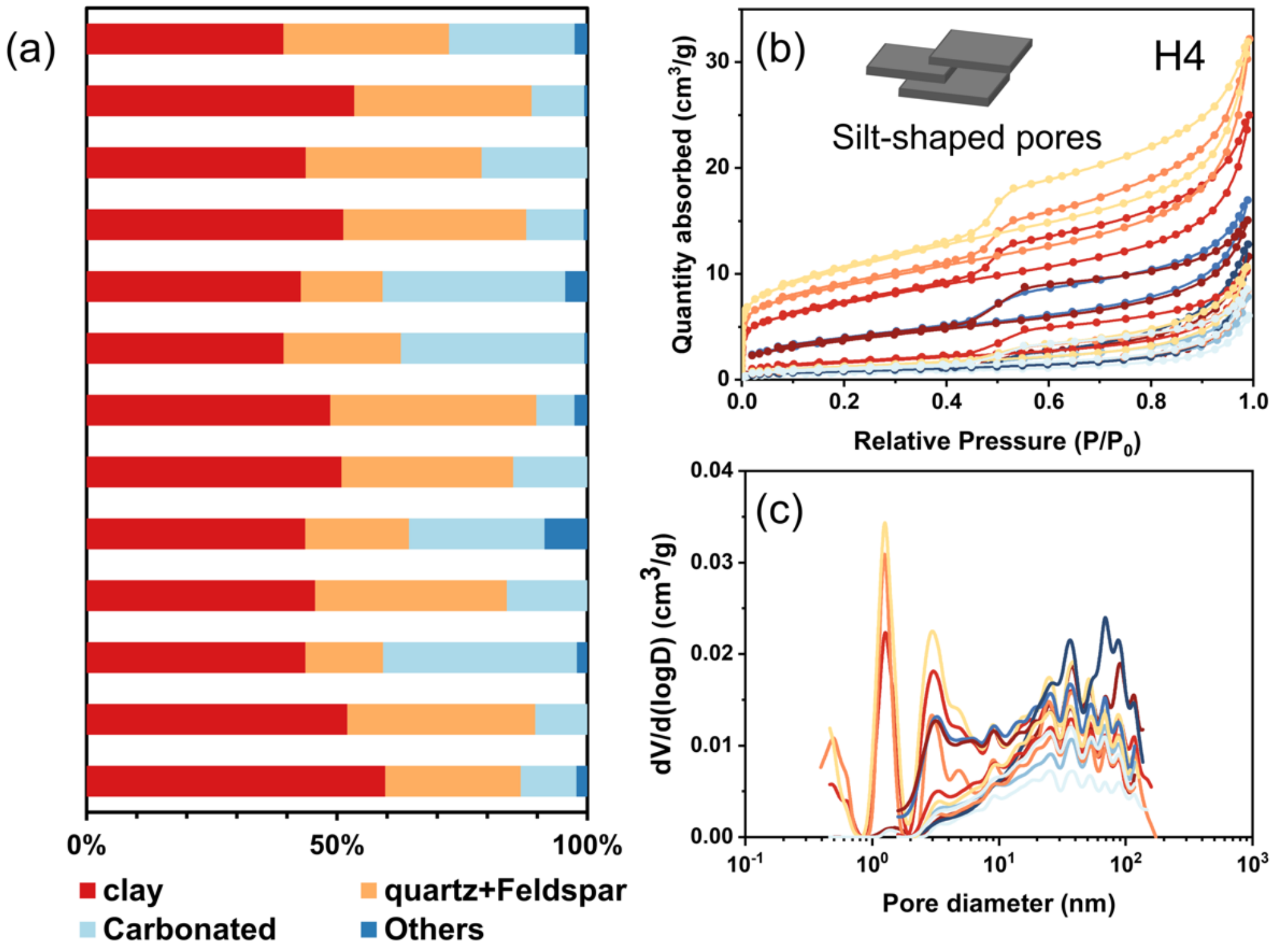

*Fig. S1 Characteristics of mineral distribution and pore structure in typical lacustrine shale (here exemplified by shale from the Songliao Basin in China): (a) distribution of mineral mass ratio based on whole-rock X-ray diffraction (XRD); (b) adsorption-desorption curves obtained from low-temperature nitrogen adsorption experiments; (c) pore size distribution characteristics based on nitrogen adsorption experiments.*

*Table S1*

*Parameters and loading numbers of mixed-component hydrocarbons.*

| Hydrocarbon Components | Name | Molecular formula | Molecular weight (g/mol) | Molecules Loading numbers | Mass ratio (%) |
|---|---|---|---|---|---|
| Light Saturates | n-octane | $C_8H_{18}$ | 114.232 | 16 | 11.02 |
| | n-nonane | $C_9H_{20}$ | 128.255 | 29 | |
| | n-decane | $C_{10}H_{22}$ | 142.282 | 41 | |
| | n-undecane | $C_{11}H_{24}$ | 156.308 | 46 | |
| | n-dodecane | $C_{12}H_{26}$ | 170.33 | 47 | |
| Heavy Saturates | n-tridecane | $C_{13}H_{28}$ | 184.361 | 45 | 50.17 |
| | n-tetradecane | $C_{14}H_{30}$ | 198.388 | 44 | |
| | n-pentadecane | $C_{15}H_{32}$ | 212.415 | 41 | |
| | n-hexadecane | $C_{16}H_{34}$ | 226.448 | 37 | |
| | n-heptadecane | $C_{17}H_{36}$ | 240.468 | 35 | |
| | n-octadecane | $C_{18}H_{38}$ | 254.502 | 60 | |
| | n-nonadecane | $C_{19}H_{40}$ | 268.521 | 31 | |
| | n-eicosane | $C_{20}H_{42}$ | 282.556 | 26 | |
| | n-heneicosane | $C_{21}H_{44}$ | 296.574 | 25 | |
| | n-docosane | $C_{22}H_{46}$ | 310.601 | 20 | |
| | n-tricosane | $C_{23}H_{48}$ | 324.627 | 20 | |
| | n-tetracosane | $C_{24}H_{50}$ | 338.654 | 16 | |
| | n-pentacosane | $C_{25}H_{52}$ | 352.68 | 15 | |
| | n-hexacosane | $C_{26}H_{54}$ | 366.707 | 12 | |
| | n-heptacosane | $C_{27}H_{56}$ | 380.733 | 11 | |
| | n-octacosane | $C_{28}H_{58}$ | 394.76 | 8 | |
| | n-nonacosane | $C_{29}H_{60}$ | 408.787 | 7 | |
| | n-triacontane | $C_{30}H_{62}$ | 422.813 | 5 | |
| Aromatics | benzene | $C_6H_6$ | 78.114 | 361 | 17.53 |
| | naphthalene | $C_{10}H_8$ | 128.174 | 110 | |
| Heteroatomic Hydrocarbons | n,n-dimethyldodecylamine | $C_{14}H_{31}N$ | 213.4 | 66 | 13.62 |
| | n-octadecanoic acid | $C_{18}H_{36}O_2$ | 284.484 | 66 | |
| Asphaltenes | ASNP | $C_{62}H_{85}$ | 847.383 | 11 | 7.66 |
| | ASPO | $C_{60}H_{81}NS$ | 830.362 | 11 | |

*Table S2*

*The composition of crude oil extracted from shales of different maturity levels in the Songliao Basin of China.*

| $R_o$ (%) | Saturates (%) | Aromatics (%) | Heteroatomic hydrocarbons (%) | Asphaltenes (%) |
|---|---|---|---|---|
| 0.7 | 58 | 19 | 20 | 3 |
| 0.9 | 55 | 21 | 20 | 4 |
| 1.1 | 62 | 18 | 17 | 3 |
| This study | 61.19 | 17.53 | 13.62 | 7.66 |

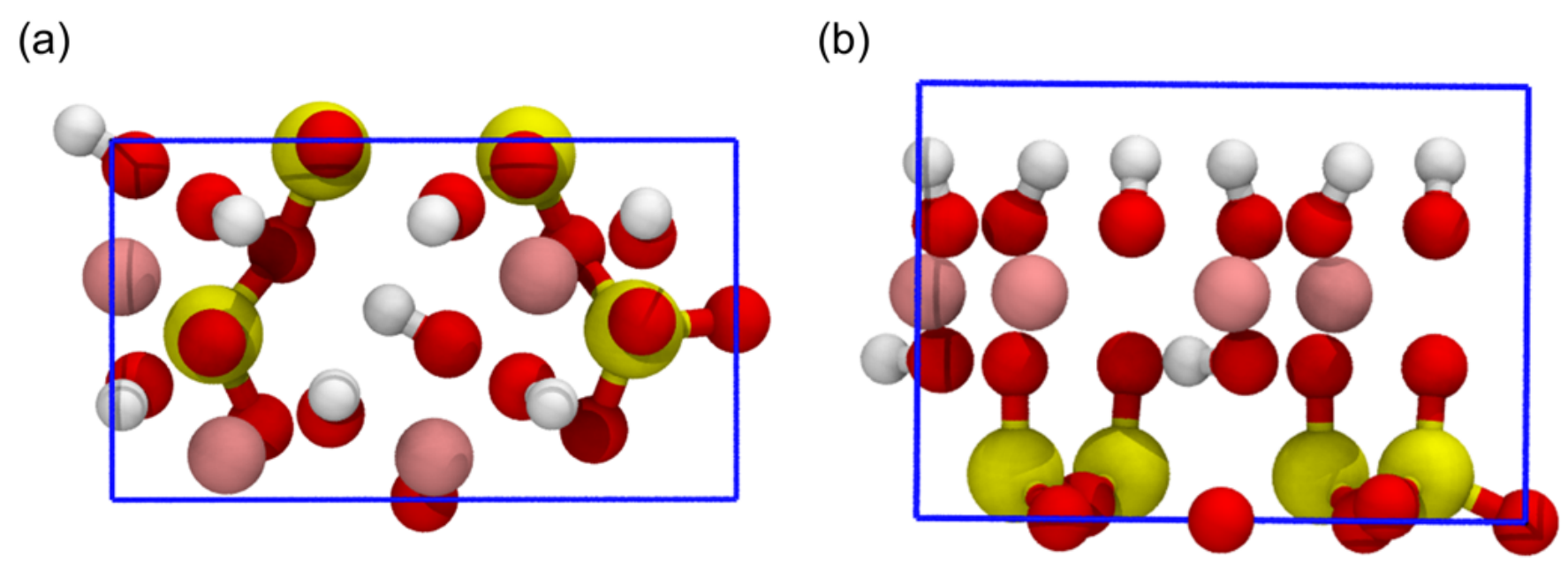

*Fig. S2 Unit cell of Kaolinite (a) top and (b) side view. Colours are: Si – yellow, Al – pink, O – red and H – white.*

*Table S3*

*Energy terms and application of ClayFF force field (Cygan et al., 2004).*

| Energy term | Potential term | Equation | Interacting atoms | Applied to |
|---|---|---|---|---|
| Total | Total | $\mathrm{E}_{total} = E_{bonded} + \mathrm{E}_{non-bonded}$ | Whole system | Whole system |
| Van der Waals | Lennard-Jones (12-6) | $E_{VDW} = \sum_{i \neq j} D_{0,i,j}\left[\left(\frac{R_{0,i,j}}{r_{ij}}\right)^{12} - 2\left(\frac{R_{0,i,j}}{r_{ij}}\right)^{6}\right]$ | 2 | Any 2 non-bonded atoms |
| Coulombic | Coulombic | $E_{Coul} = \frac{e^2}{4\pi\varepsilon_0}\sum_{i \neq j}\frac{q_i q_j}{r_{ij}}$ | 2 | Any 2 non-bonded atoms |
| Bond stretch | Harmonic | $E_{bond} = k_1\left(r_{ij} - r_0\right)^2$ | 2 | O-H bond in clay |

*Table S4*

*Energy terms and application of Charmm36 force field (Bjelkmar et al., 2010).*

| Energy term | Potential term | Equation | Interacting atoms | Applied to |
|---|---|---|---|---|
| Total | Total | $\mathrm{E}_{total} = E_{bonded} + \mathrm{E}_{non-bonded}$ | Whole system | Whole system |
| Van der Waals | Lennard-Jones (12-6) | $E_{VDW} = \sum_{i \neq j} D_{0,i,j}\left[\left(\frac{R_{0,i,j}}{r_{ij}}\right)^{12} - 2\left(\frac{R_{0,i,j}}{r_{ij}}\right)^{6}\right]$ | 2 | Any 2 non-bonded atoms |
| Coulombic | Coulombic | $E_{Coul} = \frac{e^2}{4\pi\varepsilon_0}\sum_{i \neq j}\frac{q_i q_j}{r_{ij}}$ | 2 | Any 2 non-bonded atoms |
| Bond stretch | Harmonic | $E_{bond} = k_1\left(r_{ij} - r_0\right)^2$ | 2 | Organic hydrocarbons |
| Angle bend | Harmonic | $E_{angle} = k_2\left(\theta_{ij} - \theta_0\right)^2$ | 3 | Organic hydrocarbons |
| UB | Harmonic | $E_{UB} = k_3\left(b^{1-3} - b^{1-3,0}\right)^2$ | 3 | Organic hydrocarbons |
| Dihedral | Periodic function | $E_{dihedral} = k_4\left[1 - \cos\left(n\varphi - \delta\right)\right]$ | 4 | Organic hydrocarbons |
| Improper | Harmonic | $E_{improper} = k_5\left(\omega + \omega_0\right)^2$ | 4 | Aromatics hydrocarbons; Asphaltenes |
| CMAP | Energy Table | $E_{CMAP} = \upsilon_{CMAP}\left(\phi, \psi\right)$ | Map | Organic hydrocarbons |

Where: $D_0$ is the depth of the potential well; $r_{ij}$ is the separation distance of atoms i and j; $\theta_{ijk}$ is the angle of atoms i, j, and k; $R_{0,i,j}$ is radius that corresponds to the minimum LJ potential energy; $q_i$, $q_j$ are partial charges of two atoms; $\varepsilon_0$ is the dielectric constant, $r_0$ and $\theta_0$ are equilibrium values of radius and angles; $k_1$–$k_5$ are prefactors; $n$ is an integer ≥ 0; $b^{1-3}$ and $b^{1-3,0}$ respectively represent the distance between atoms 1-3 and the equilibrium value; $\varphi$ is the angle of the dihedral, while $\delta$ is the angle of the phase angle; $\omega$ and $\omega_0$ respectively are the angle of the improper angle and its equilibrium value.

*Table S5*

*The functional parameters of ClayFF force field employed in the MD simulations of the kaolinite system (Cygan et al., 2004)*

| Molecular interactions | Species | Atom type | Mass (g/mol) | Charge (e) | $R_0$ (nm) | $D_0$ (kJ/mol) |
|---|---|---|---|---|---|---|
| Non-bonded interactions | Tetrahedral Si | st | 28.09 | 2.1000 | 3.302E-01 | 7.701E-06 |
| | Octahedral Al | ao | 26.98 | 1.5750 | 4.271E-01 | 5.564E-06 |
| | Hydrohyl H | ho | 1.008 | 0.4250 | 0.000E+00 | 0.000E+00 |
| | Hydroxyl O | oh | 16 | -0.9500 | 3.166E-01 | 6.502E-01 |
| | | | | | $R_0$, nm | k, kJ/mol $nm^2$ |
| Bonds | Hydroxyl bond | | | | 1.000E-01 | 4.635E+05 |

*Table S6*

*The subsequent list specifies the atomic types that have been assigned to the system's organic hydrocarbon molecules. The non-bonded and bonded parameters pertinent to each atom type are readily accessible at: http://mackerell.umaryland.edu/charmm_ff.shtml.*

| Species | Atom type |
|---|---|
| Aliphatic C for $CH_3$ | CG331 |
| Aliphatic C for $CH_2$ | CG321 |
| Alphatic proton for $CH_2$ | HGA2 |
| Alphatic proton for $CH_3$ | HGA3 |
| Aromatic C | CG2R61 |
| Aromatic H | HGR61 |
| Carbonyl C for esters and carboxylic acids (neutral) | CG2O2 |
| Carbonyl O for amides, esters, carboxylic acids (neutral), aldehydes and urea | OG2D1 |
| Hydroxyl oxygen | OG311 |
| Polar H | HGP1 |
| Neutral trimethylamine nitrogen | NG301 |
| Sulphur for SH or -S- | SG311 |
| Polar H for thiol | HGP3 |
| Aliphatic C for CH with one H | CG311 |
| Alphatic proton for CH | HGA1 |
| Double bound neutral 6-mem planar ring for pyrrole and pyran | NG2R60 |

Taking stearic acid as an example, this study demonstrates the method for assessing system convergence. Fig. S3 (d) shows the RMSD curve across the simulation timeframe for stearic acid. It can be observed that the RMSD curve stabilizes after 30 ns, indicating that comprehensive exploration of all accessible phase spaces by each stearic acid molecule, thereby indicating a state of stability within the system relative to stearic acid molecules. Nonetheless, this research primarily focuses on the interactions between hydrocarbon molecules adjacent to both basal surfaces of kaolinite and the clay. Hence, it is more reasonable to assess system convergence by analyzing the evolution of linear density profiles, utilizing DynDen software. Fig. S3 (a) shows the variation in box size along the Z-axis during the simulation. The box size gradually decreases as the simulation duration increases, attributable to molecular space compression caused by pressure. Fig. S3 (b) and (c) show the density evolution of the system and stearic acid over the simulation period. Four stable adsorption layers formed on the right SI basal surface of the kaolinite at around 10 ns, indicating that hydrocarbon molecules on the non-polar surface are predominantly governed by van der Waals interactions. On the contrary, the HY surface generated four adsorption layers that were virtually stable at around 40 ns. This was attributed to the complex charge interactions and hydrogen bonds that existed between the polar surface and the oxygen-containing functional group head of stearic acid. Consequently, for heteroatomic hydrocarbons like stearic acid, the simulation system requires approximately 40 ns converge. Fig. S3 (e) and (f) present the pair correlation coefficient graphs for the overall system and the stearic acid, respectively, and their results also contribute as benchmarks for assessing local density equilibrium (Degiacomi et al., 2021).

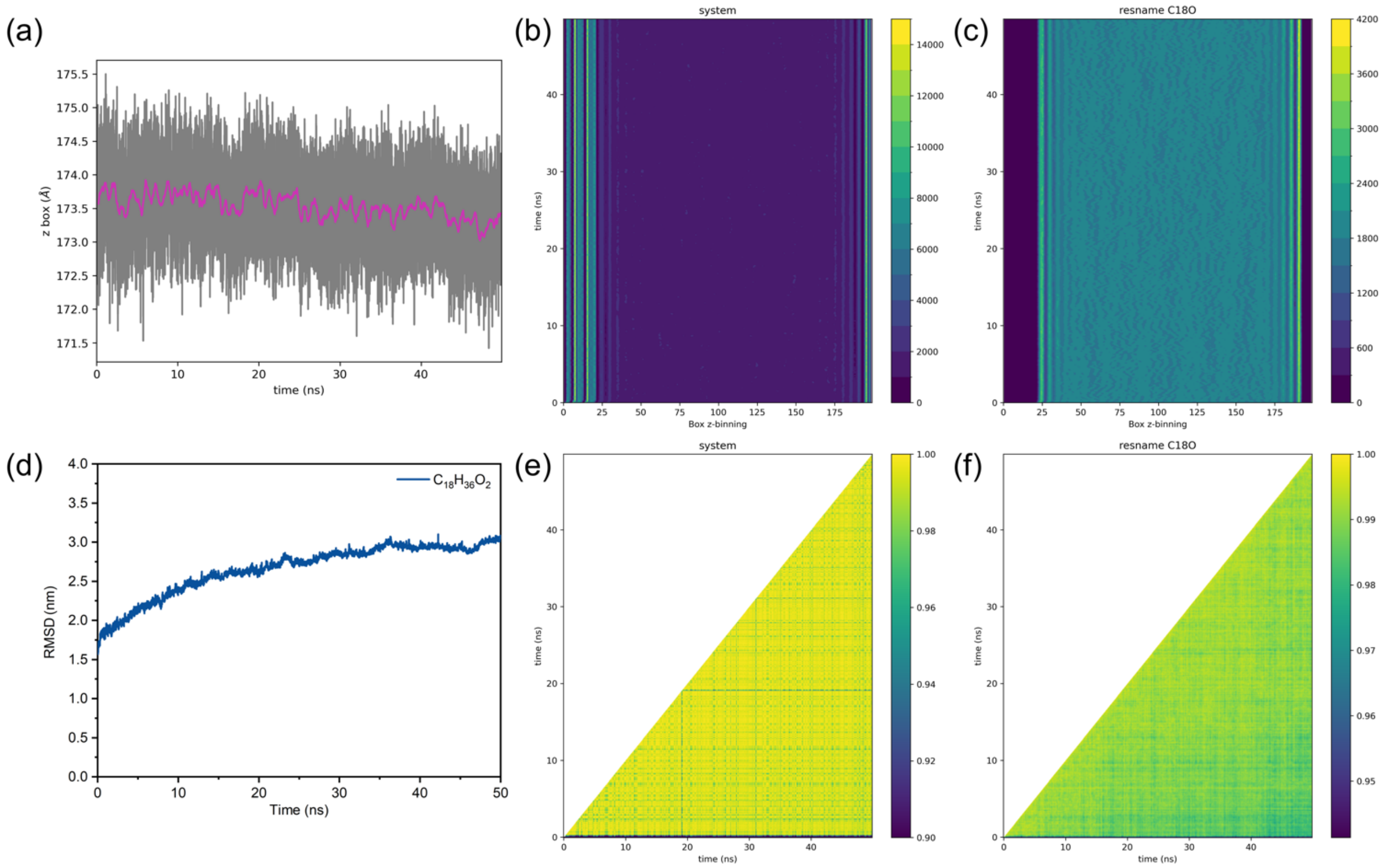

*Fig. S3 Example of convergence assessment for the isolated monomeric hydrocarbon simulation system, represented by stearic acid: (a) fluctuations in the size of the simulation box in the Z*

*direction during a 50 ns simulation period. The gray line shows the size change at each step, and the palatinate purple line represents the average size of the box; (b) and (c) show the density changes of the overall system and stearic acid over the simulation time, respectively; (d) the RMSD curve of stearic acid; (e) and (f) are the pair correlation coefficient graphs of the overall system and hydrocarbon molecules, respectively, with light-colored areas indicating high correlation.*

In multi-component oil mixture systems, global RMSD is inadequate for convergence assessment. Thus, the DynDen software was employed to analyze the density evolution of different hydrocarbon components in the mixed system over the simulation period, to evaluate the system's convergence. Fig. S4 and S5 respectively display the density alterations of various hydrocarbon components under simulation in the NPT ensemble and during periodic annealing processes. As shown in Fig. S4 (a), the system has not completely converged after 200 ns of simulation time, as variations in the number of adsorption layers of the hydrocarbon mixture on the basal surfaces of kaolinite continue throughout simulation duration. From the density evolution graphs of different hydrocarbon components, it is evident that light saturated hydrocarbons predominantly stabilize after 150 ns, mainly distributed in the pore centers and existing in the bulk fluid. In contrast, heavy saturated hydrocarbons continue to form a fifth adsorption layer at 200 ns, as depicted in Fig. S4 (c), markedly influenced by the asphaltene component. Fig. S4 (f) illustrates that the density of the asphaltene component fluctuates substantially over time, with a near-stable double-layer adsorption occurring on kaolinite's left HY basal surface after 50 ns. On the SI basal surface, characteristics of four-layer adsorption emerge at 150 ns, subsequently affecting the adsorption of heavy saturated hydrocarbons on this surface. Heteroatomic hydrocarbons stabilize on the HY surface after 150 ns, while a dynamic coexistence of adsorption and desorption persists on the SI basal surface. Aromatic hydrocarbon components maintain relative stability on the HY basal surface during the simulation, exerting minimal impact on the adsorption of other molecules on the SI surface. In summary, traditional simulation methodologies encounter difficulties in achieving convergence when applied to systems of this complexity. Hydrocarbon molecules from various components are easily impacted by competitive adsorption and interactions with clay basal surfaces during the simulation. They become trapped in local energy minima, which hinders their ability to escape and creates a misleading representation of fluid behavior.

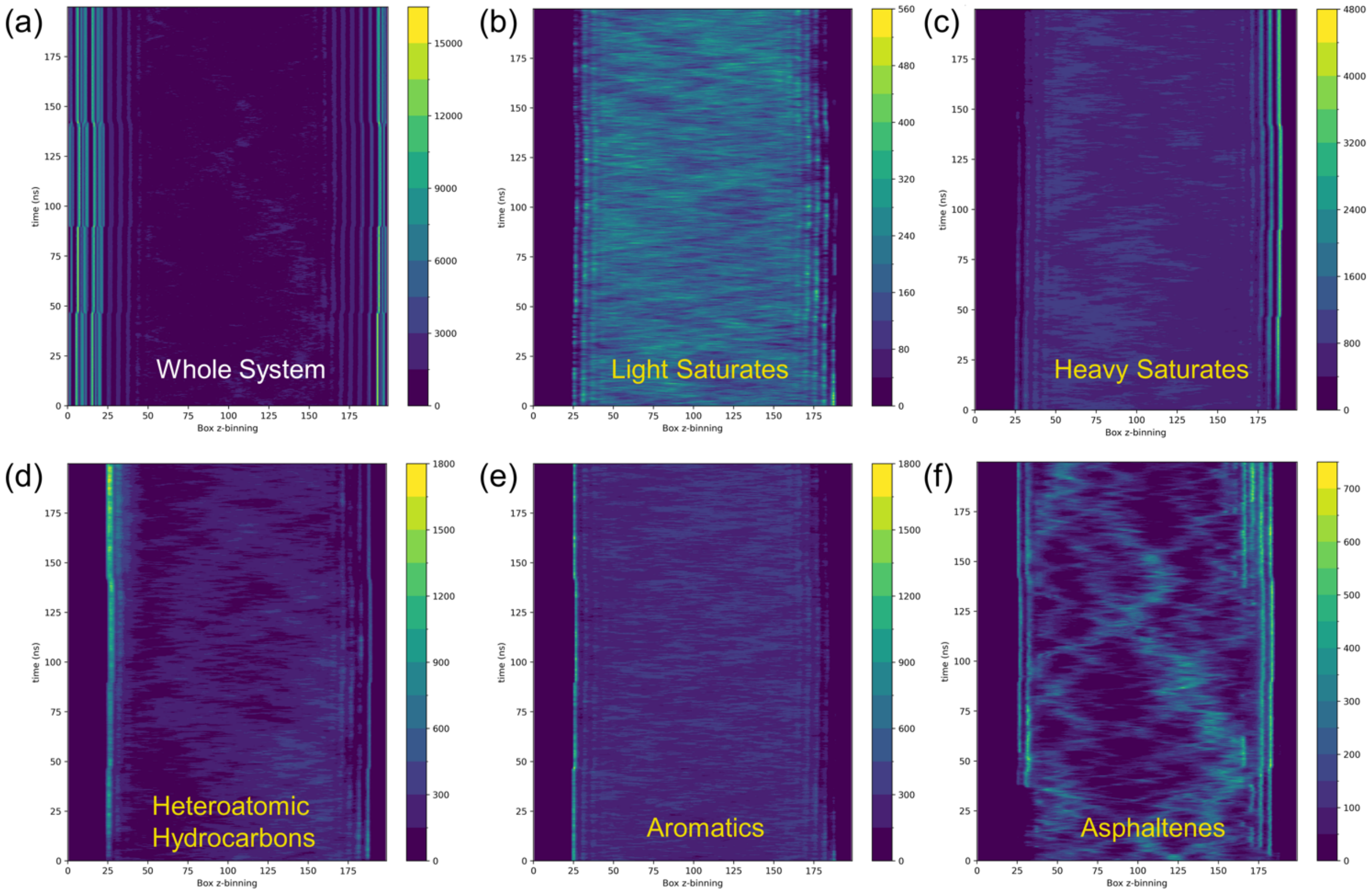

*Fig. S4 The evolution of system density and the density of each hydrocarbon component over time in simulations under NPT ensemble conditions, where light colors represent areas of high density, and conversely, dark colors represent areas of low density.*

Fig. S5 shows the density evolution of different components within mixed system during a periodic annealing process. The evolution of the system's density indicates a gradual convergence subsequent to the temperature reduction to 348 K. As illustrated in Fig. S5 (a), a total of three adsorption layers are finally formed on the left HY basal surface of kaolinite, whereas four adsorption layers are formed on the right SI surface. Compared to simulations under the NPT ensemble, the density evolution of different hydrocarbon components experiences marked variations. Fig. S5 (b) indicates that the alterations in the light saturated hydrocarbon component are relatively minor, with a predominant distribution in the bulk fluid, although some adsorption on kaolinite's SI basal surface is observed. Heavy saturated hydrocarbons exhibit negligible adsorption on the HY basal surface, predominantly due to competitive adsorption by heteroatomic hydrocarbons. Fig. S5 (d) reveals that heteroatomic hydrocarbons are fully adsorbed on the HY surface and maintain stability throughout the annealing process, suggesting complete convergence for this component. Heavy saturated hydrocarbons primarily adsorb on the SI basal surface of kaolinite, where they sustain a stable four-layer adsorption for a duration of 150 ns. The aromatic hydrocarbons maintain essential stability during the annealing, predominantly adsorbing on the HY basal surface's 1$^{st}$ adsorption layer and competing with heteroatomic hydrocarbons. Compared to standard simulations, the variations in asphaltenes during the annealing process are more pronounced. As illustrated in Fig. S5 (f), asphaltenes initially demonstrate double-layer adsorption on the HY surface. With rising temperatures, this surface becomes predominantly occupied by aromatic and heteroatomic hydrocarbons, prompting the desorption of asphaltene molecules into the bulk fluid. As the temperature decreases to 348 K, asphaltene molecules progressively establish a four-layer adsorption on the SI surface, engaging in strong competition with heavy saturated hydrocarbons. The density evolution attains stability after 150 ns, indicating convergence for the asphaltene component. These analyses demonstrate that dynamic density assessment via DynDen software can effectively evaluate the convergence of complex mixed systems, ensuring the representativeness of simulation outcomes.

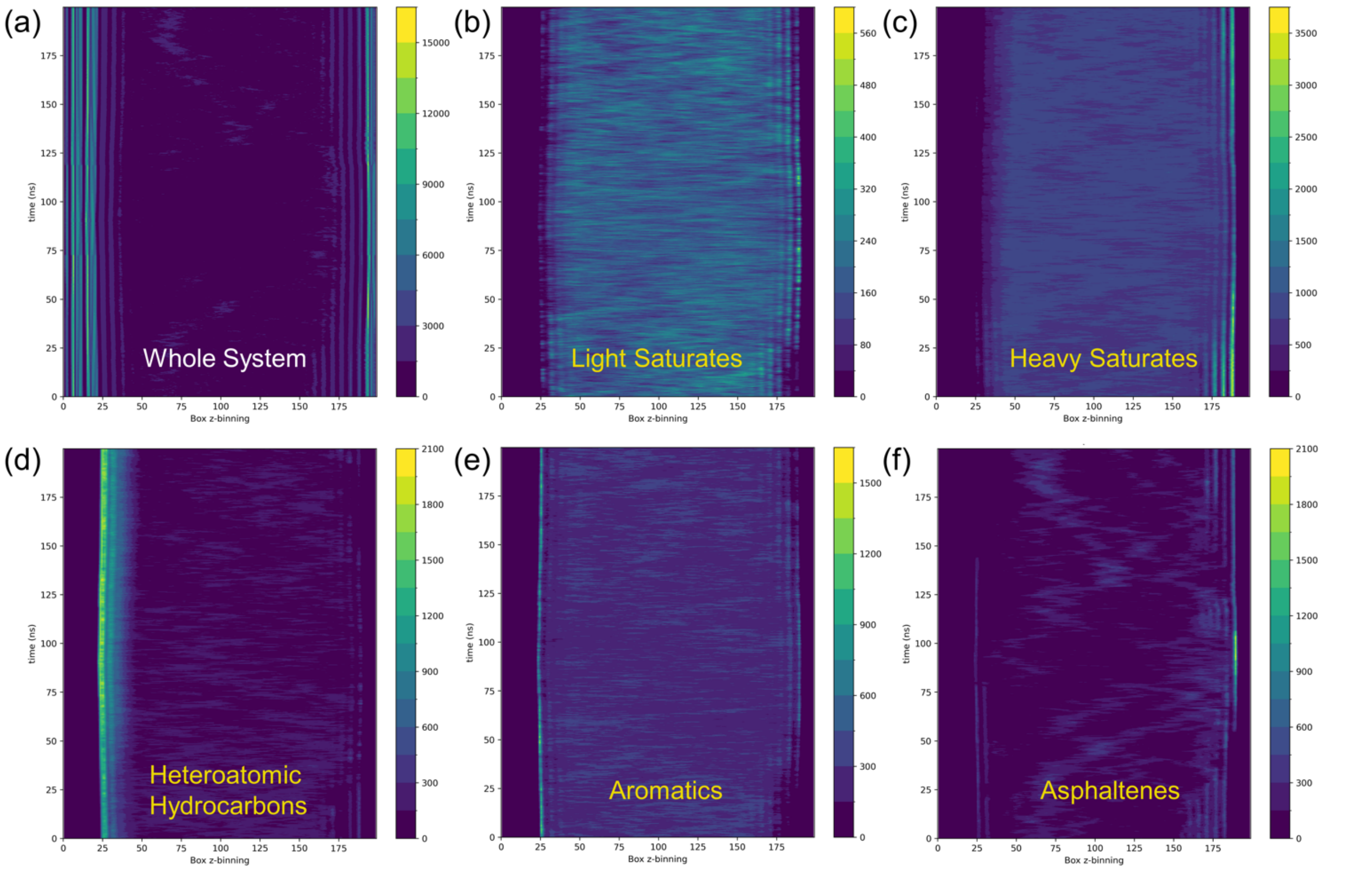

*Fig. S5 The evolution of system density and the density of each hydrocarbon component over time in periodic annealing simulations, where light colors indicate areas of high density, and conversely, dark colors indicate areas of low density.*